\documentclass[preprint,12pt]{elsarticle}

\usepackage[export]{adjustbox}
\usepackage{color}
\usepackage{enumerate}
\usepackage{enumitem}
\usepackage{float}
\usepackage{graphicx}
\usepackage{listings}
\usepackage{multirow}
\usepackage{subcaption}
\usepackage{url}
\usepackage[table]{xcolor}
\usepackage{tabularx} 

\graphicspath{{figures/}}
\newtheorem{pattern}{Pattern}

\journal{Computers in Industry}

\begin{document}

\begin{frontmatter}


\title{A Reference Model and Patterns for Production Event Data Enrichment}


\author[1,3]{M.C.A. van der Pas\corref{cor1}}
\ead{m.c.a.v.d.pas@tue.nl}

\author[2]{A.E. Ak\c{c}ay}

\author[1]{R.M. Dijkman}

\author[1]{I.J.B.F. Adan}

\author[3]{J. Walker}

\cortext[cor1]{Corresponding author}

\affiliation[1]{organization={Department of Industrial Engineering \& Innovation Sciences, Eindhoven University of Technology},
            addressline={PO Box 513}, 
            city={Eindhoven},
            postcode={5600MB},
            country={The Netherlands}}

\affiliation[2]{organization={Department of Mechanical and Industrial Engineering, Northeastern University},
            addressline={360 Huntington Ave.},
            city={Boston, MA},
            country={United States}}

\affiliation[3]{organization={Semaku B.V.},
            addressline={Torenallee 20},
            city={Eindhoven},
            postcode={5617BC},
            country={The Netherlands}}

\begin{abstract}
With the advent of digital transformation, organisations are increasingly generating large volumes of data through the execution of various processes across disparate systems. By integrating and aggregating data from these heterogeneous sources, it becomes possible to derive new insights essential for tasks such as monitoring and analysing process performance. Typically, this information is extracted during a data pre-processing or engineering phase. However, this step is often performed in an ad-hoc manner and is time-consuming and labour-intensive.
To streamline this process, we introduce a reference model and a collection of patterns designed to enrich production event data.
The reference model provides a standard way to store and extract production event data. The patterns describe common data aggregation and information extraction tasks. The reference model is constructed by combining the ISA-95 industry standard with the Event Knowledge Graph formalism. The patterns are developed based on empirical observations from event data sets originating in manufacturing processes and are formalised using the reference model.
The ability of the proposed patterns to accelerate data aggregation and facilitate knowledge exchange has been validated through an empirical evaluation involving manufacturing professionals from different roles and production environments.

\end{abstract}


\begin{highlights}
    \item Collected generic patterns for aggregating manufacturing event data;
    \item Defined a reference model that describes the terminology;
    \item Formalised the patterns and illustrated their applications;
    \item Shared reference implementation and example data in a public repository;
    \item Evaluated the approach with professionals.
\end{highlights}


\end{frontmatter}

\section{Introduction}
\label{sec:introduction}
In large manufacturing organisations, numerous heterogeneous information systems are deployed to record data about the execution and performance of operational processes. With digital transformation and the Industrial Internet of Things, the number of systems and the volume of data are increasing \cite{Afrin2025IndustrialIndustries,Lee2021UnderstandingDiscovery}. The data come in many different forms, ranging from master data describing a company's assets to quality measurements of products or services delivered to a customer. Additionally, there is substantial variation in sources, for example, a human manually entering a text or numeric value in a system or the fully automated capturing of high-frequency sensor readings. 

However, there is often a big gap between the raw data captured by the information systems and the relevant insights, or information, useful to a company~\cite{Rowley2007TheHierarchy}. Cleaning the data and deriving the `hidden' information is commonly done in a data pre-processing or data engineering activity, which can be challenging and time-consuming, as it typically requires ad hoc and manual effort \cite{Li2025DataStrategy,Dogan2021MachineManufacturing}. It also requires domain knowledge, especially when analysing event data \cite{Schuster2022UtilizingReview}.


To address this challenge, we introduce a set of patterns to derive `hidden' and aggregated information from data. Patterns are common solutions to frequent problems~\cite{Alexander1977AConstruction}; in our context, they address the challenge of extracting information from event data.
While patterns have been successfully adopted in various domains, most notably in software design by the `Gang of Four' \cite{Gamma1994DesignSoftware} and Fowler \cite{Fowler2002PatternsArchitecture}, they are also highly relevant to the business process community through workflow \cite{VanderAalst2003WorkflowPatterns}  and event log patterns \cite{Suriadi2017EventLogs}.
The patterns facilitate more efficient information extraction and serve as a foundation for automation. 
Because the patterns described in our work are inspired by practical use cases, we adopt the definition by Fowler 
\cite{Fowler1997AnalysisModels}: ``A pattern is an idea that has been useful in one practical context and will probably be useful in others." 
Ultimately, our work aims to accelerate data aggregation and promote the systematic retention of knowledge on how to extract information.
This paper introduces, to the best of our knowledge, the first structured collection of reference-model-based patterns for enriching production event data in manufacturing environments.

We derive the patterns from observations about data sets from four different manufacturing environments. To describe and formalise the patterns, we define the reference model. The reference model combines the well-known standard ISA-95 \cite{OPCFoundation2019OPCReference} and the Event Knowledge Graph formalism \cite{Esser2021Multi-DimensionalDatabases,Fahland2022ProcessGraphs}. The benefit of using this reference model is that it both provides a uniform terminology to describe the patterns and serves as a basis for extracting information from databases that are structured according to those standards. Furthermore, reference models and ontologies are commonly used for knowledge reuse \cite{Guo2024AnProducts}. Subsequently, we define each of the patterns in terms of the reference model and motivate them by providing use cases. We also identify which patterns were observed in the four industrial production environments studied and evaluate the approach with professionals from multiple manufacturing environments, covering both IT- and business-oriented roles.



The remainder of this work is structured as follows. First, in Section~\ref{sec:related_work} we present related work on patterns and enrichment of event data. Section~\ref{sec:methodology} gives an overview of the approach we used to arrive at the presented patterns. Subsequently, we present the reference model in Section~\ref{sec:reference_model} followed by a description of the patterns in Section~\ref{sec:production_trace_patterns}. Finally, we provide an empirical evaluation of the relevance with professionals working from industry in Section~\ref{sec:evaluation} and concluding discussion in Section~\ref{sec:conclusion}.

\section{Related work}
\label{sec:related_work}
In this section, we present previous work related to patterns for deriving insights from data, specifically data originating from manufacturing processes. First, we give some background information on the usage of patterns. Subsequently, we discuss the work related to the application of patterns on event data in general. We position our work within the event processing and reasoning literature, where the generation of new insights from event data is also of interest. Finally, we discuss several ontologies for the manufacturing domain. Those ontologies describe common data patterns in manufacturing environments and provide the basis for the concepts used in the patterns presented in this work.

Suriadi et al. \cite{Suriadi2017EventLogs} describe patterns to detect imperfections in event data. The proposed patterns can be used in the first place to detect common data quality issues in event data sets and subsequently remove them. Those patterns are therefore also part of the data pre-processing step. Based on experts' knowledge and common sense, it is also possible to define inference rules that can be used to derive missing data (identifiers) from an incomplete event data set \cite{Swevels2023InferringGraphs}. Another approach is to use process models to infer missing or unobserved events \cite{Fahland2021InferringQueues}. In addition to defining patterns or rules with the help of experts, it is also possible to learn operators that can be used to complete event data \cite{Muller2016CaseReasoning}. Lee et al. \cite{Lee2008RFID-basedChain} propose a model and a system to combine event data with information on the bill of material to reconstruct the production trace of products. In contrast to the approaches presented above, the patterns proposed in this work are not targeted to detect imperfections in the data, but rather to enrich the data with initially hidden information.

Several works propose event processing and stream reasoning methods to enrich event data. Anicic et al. \cite{Anicic2011EP-SPARQL:Reasoning} propose a graph query language for event processing, which can also be used to express patterns on top of multi-dimensional event data. Park and van der Aalst \cite{Park2025OperationalApproach} also look at multi-dimensional event data and define metrics for monitoring business processes. Bonte et al. \cite{Bonte2018StreamingStreams} propose a reasoning approach for complex event processing to derive information from an event stream annotated with semantics from an ontology \cite{Tommasini2017TowardsProcessing}. Teymourian \cite{Teymourian2014AProcessing} presents an extensive study of use cases and approaches for knowledge-based event processing, including the enrichment of event streams. Those works show the relevance of deriving information from event data and propose several general methods to retrieve new information, but do not provide the patterns to retrieve domain-specific information. We contribute to this literature stream by describing a set of patterns specifically for the manufacturing domain.

Within the manufacturing domain, there have been several efforts to model knowledge and enable data access using ontologies. Some of the first ontologies were proposed by Lemaignan et al. \cite{Lemaignan2006MASON:Domain} and Borgo et al. \cite{Borgo2007FoundationsManufacturing}. More recently, with the emergence of Industry 4.0, sensors have become an important concept \cite{Giustozzi2018ContextProposal,Alvanou2018AnAnalytics}. Furthermore, there are efforts to develop a foundational ontology for the industrial manufacturing domain \cite{Kulvatunyou2018TheProject} and ontologies for specific manufacturing settings \cite{Cao2022AManagement}. Yang et al. \cite{Yang2023Ontology-basedWorkflow} build on previous work to define an ontology for industrial production workflows. Byun et al. \cite{Byun2017EfficientEvents} propose a graph-based model for describing events for traceability in manufacturing and the supply chain. Previous work defining manufacturing ontologies provides an overview of the data structures and information that is relevant in manufacturing environments. However, they do not describe how to derive the information, which is what we aim to achieve with the patterns collected in this work.

\section{Methodology}
\label{sec:methodology}
We identified the patterns based on observations about existing data sets and subsequently formalised them using a reference model that was constructed based on existing standards. This section first explains how the reference model was developed based on existing standards, and then how patterns were extracted from the data sets.

\subsection{Reference Model}
The reference model presented in Section~\ref{sec:reference_model} is a consolidation of a general model and relevant ontologies from the manufacturing domain (Section~\ref{sec:related_work}). General concepts from the Event Knowledge Graph (EKG) model \cite{Esser2021Multi-DimensionalDatabases,Fahland2022ProcessGraphs} are used as the basis of our model. EKG is proposed as a graph-based solution to deal with multi-dimensional event data and can be applied to any event log, including events generated from manufacturing processes \cite{Swevels2023InferringGraphs}.

As the production trace patterns target a manufacturing environment, we included more detailed manufacturing entities in the model, which are in line with the existing ontologies referred to in Section~\ref{sec:related_work}. In addition to general concepts, more detailed concepts are included in the model. Those detailed concepts are derived from the data sets and are used to motivate and describe the production trace patterns in more detail. The exact data structures, data sources, and terminology will differ between organisations, but the concepts can be used as inspiration and a guideline for applying the patterns.

\begin{figure}[t]
    \centering
    \includegraphics[width=\linewidth]{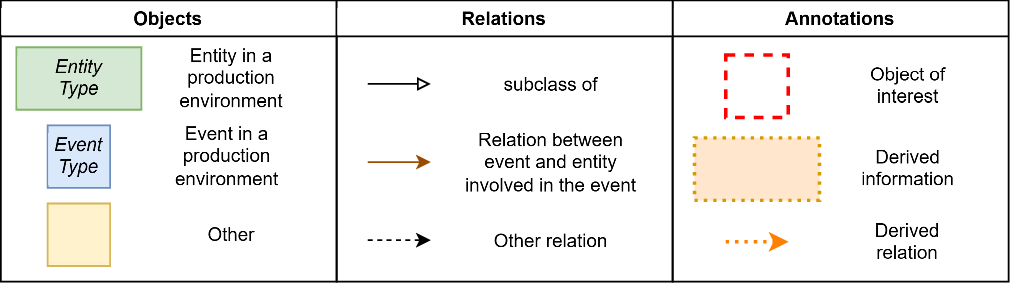}
    \caption{Notation that is used to define the reference model and the pattern examples.}
    \label{fig:legend}
\end{figure}

Figure~\ref{fig:legend} gives an overview of the notation used to define the reference model and the pattern examples. The main objects are the events (blue) and entities (green); other objects are represented in yellow. In addition, we differentiate various types of relationships between objects. In the taxonomy, the subclass relation is used and the relation between event and entity is important. Other relations are represented using a dashed line. Finally, several annotations are used in the examples of the patterns. In those examples, the objects of interest are marked with a red dashed outline and the derived information or relations are represented in orange.

Note that the main goal of the model presented in this work is to reach a common understanding of the manufacturing concepts that are used to define and describe the production trace patterns. Therefore, our goal is to define a reference model and not a comprehensive ontology for the manufacturing domain.

\subsection{Patterns}
\label{sec:patterns_data_sets}
The data sets from several (discrete) manufacturing companies in different domains inspired the patterns described in this work. The data sets contain production traces, defined as events and associated entities illustrating the production process of items (e.g., batches or lots). Although the use cases for the patterns presented in this work are scoped to the production trace in one organisation (internal traceability \cite{Schuitemaker2020ProductReview}), the same concept can be applied to a group of organisations within a supply chain. Production traces may include contextual data that describe events, entities, and their interrelations.

The first data set comes from the semiconductor back-end operations, handling the assembly and testing steps in a batch process flow shop \cite{Lin2015SimulationManufacturing}. The second data set is from the automotive domain, specifically a shop floor simulation where AGVs (Automated Guided Vehicles) transport components between workstations. For reference, an anonymised version of this data set is published on GitHub\footnote{\url{https://github.com/gitmpje/production-trace-patterns/tree/6e4581633b05f43b019773706972d126afa1d682/use_cases/automotive}}. Another data set comes from a contract manufacturing company which produces to order with diverse equipment and human-driven operations (e.g., welding). The final set is from an industrial automation equipment manufacturer and details an assembly line where multiple workstations are managed by operators. The process includes several quality inspections that take place during and at the end of the process.

Table~\ref{tab:overview_data_sets} gives an overview of the data sets, describing the type of manufacturing setting, the level of automation, the level of digitisation (i.e., to what extent the data are automatically collected and communicated using digital systems), and the ISA-95 \cite{OPCFoundation2019OPCReference} level (at which event data are collected). The levels considered are:
\begin{itemize}
    \item[level 3:] Data from MES or systems on the same control/activity level.
    \item[level 2:] Data registered at a resource (e.g. machine) level.
    \item[level 1:] Data concerning sensing of the production process.
\end{itemize}

The use cases that led to the patterns were derived from the data sets and discussions with practitioners. Subsequently, the patterns observed in the use cases were generalised to arrive at the patterns that we present in this work.

\begin{table}[t]
\caption{Overview of the manufacturing settings that provided inspiration and motivation for the production trace patterns.}
\label{tab:overview_data_sets}
\begin{tabular}{c|c|c|c|c}
  \textbf{Data set} & \textbf{\begin{tabular}[c]{@{}c@{}}Manufacturing\\ setting\end{tabular}} & \textbf{\begin{tabular}[c]{@{}c@{}}Level of\\ automation\end{tabular}} & \textbf{\begin{tabular}[c]{@{}c@{}}Level of\\ digitization\end{tabular}} &  \textbf{\begin{tabular}[c]{@{}c@{}}ISA-95\\ level\end{tabular}} \\ \hline \hline
  \begin{tabular}[c]{@{}c@{}}Semi-conductor\\ back-end\end{tabular} & Flow Shop & high & high & 1-3 \\ \hline
  Automotive & Job Shop & high & high & 2-3 \\ \hline
  \begin{tabular}[c]{@{}c@{}}Contract\\ manufacturing\end{tabular} & Job Shop & low & low & 2-3 \\ \hline
  \begin{tabular}[c]{@{}c@{}}Industrial\\ equipment\end{tabular} & Assembly Line & low & high & 1-3 \\ \hline
\end{tabular}
\end{table}

\section{Reference Model}
\label{sec:reference_model}
\begin{figure}
    \centering
    \includegraphics[width=\linewidth]{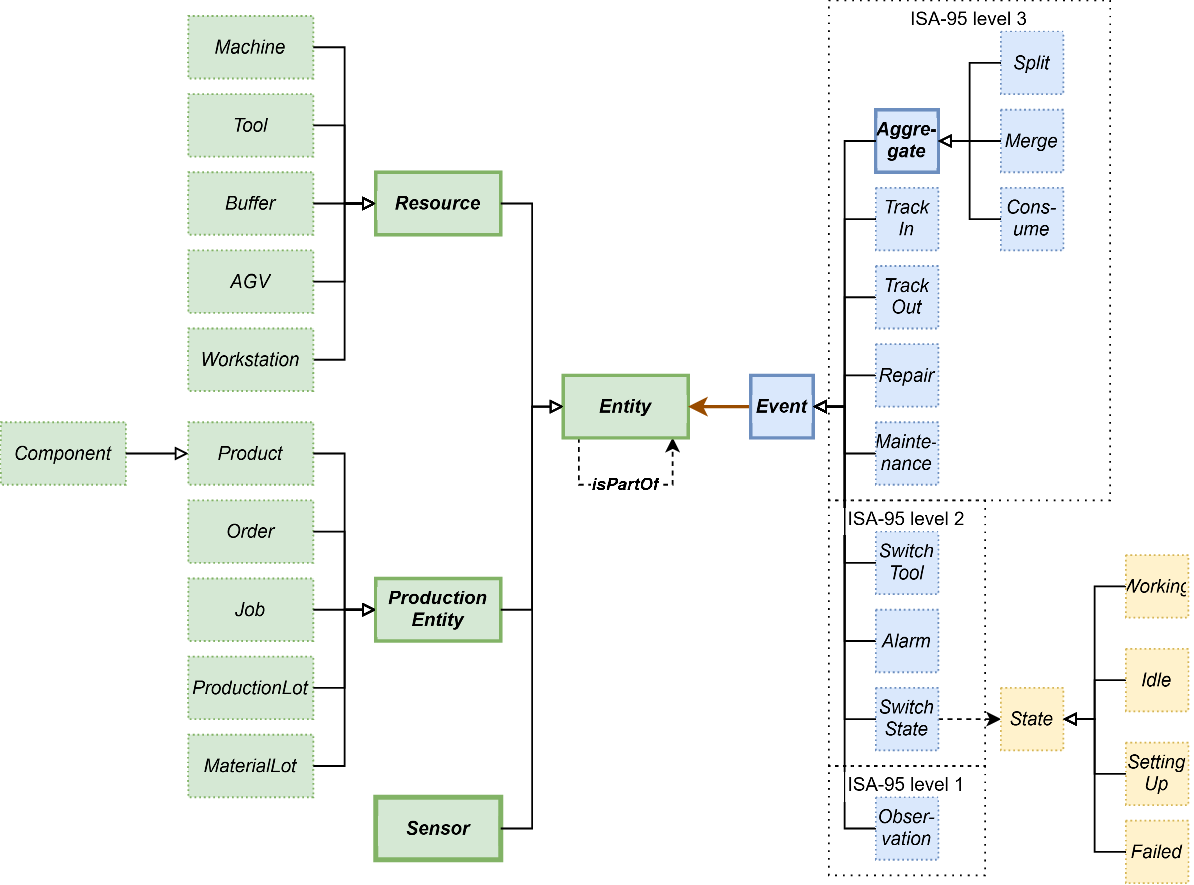}
    \caption{Reference model describing the different event and entity types, and their relations, that are used to describe the production trace patterns. The concepts in \textit{\textbf{bold}} define the high-level model.}
    \label{fig:reference_model}
\end{figure}

In this section we describe the main concepts of the reference model that is used to define the patterns. The concepts are based on existing definitions and standards. Therefore, the model also serves as a basis for extracting information based on the patterns when databases are structured according to those concepts. An overview of our model can be found in Figure~\ref{fig:reference_model}.

The high-level model covers the concepts that are at the core of business processes and their execution. In business processes typically two types of entities can be distinguished. On the one hand, there are resources that execute a (production) process, like machines and operators, and on the other hand, there are entities that are the subject of a (production) process, for example the products and orders. Additionally, sensors are important in observing and monitoring today's processes. Those concepts, or variants thereof, can also be found in manufacturing environments \cite{Lemaignan2006MASON:Domain,Borgo2007FoundationsManufacturing,W.Long2008ConstructOWL,Usman2013TowardsOntology,Giustozzi2018ContextProposal}. A subset of concepts is selected such that all entities used in the patterns are represented. The subset consists of the following concepts, which are subclasses of \texttt{Entity}:
\begin{itemize}
    \item \texttt{Resource}: resources execute the production processes by performing manufacturing operations \cite{Lemaignan2006MASON:Domain,Borgo2007FoundationsManufacturing,W.Long2008ConstructOWL,Usman2013TowardsOntology};
    \item \texttt{ProductionEntity}: other entities involved in the production process, like components, products \cite{Usman2013TowardsOntology} and orders \cite{Borgo2007FoundationsManufacturing};
    \item \texttt{Sensor}: sensors are used to measure/observe processes, which is important in Industry 4.0 manufacturing environments \cite{Giustozzi2018ContextProposal}.
\end{itemize}
The (manufacturing)process \cite{Giustozzi2018ContextProposal,Usman2013TowardsOntology} is not included in the model, but the \texttt{Event}s are a result of the processes executed by resources on production entities. Those five concepts (\texttt{Event}, \texttt{Entity}, \texttt{Resource}, \texttt{ProductionEntity}, and \texttt{Sensor}) define the high-level model, together with the \texttt{entity} relation between \texttt{Event} and \texttt{Entity}. In the model, we do not include the specific attributes that an object can have. However, an \texttt{Event} should at least have a timestamp at which the event occurred. A description of the more specific concepts can be found in the GitHub repository\footnote{\url{https://github.com/gitmpje/production-trace-patterns/blob/main/production-trace-patterns.ttl}}. Note that the more specific concepts are a collection of entities and events that are of interest in the data sets and therefore not complete.

Next to the taxonomy of entities and events, several patterns leverage the notion that a certain entity is either logically or physically included in another entity. For example, a component can be \textit{part of} a product, and a product can in turn be \textit{part of} a production lot. This is a very common relation and is known as the \texttt{isPartOf} (or `part-whole') design pattern \cite{Gerstl1996AApplications,OntologyDesignPatternsODP2010Submissions:PartOfOdp,DublinCore2023DCMI:Of}.

\section{Production trace patterns}
\label{sec:production_trace_patterns}
In this section, we present the production trace patterns. Inspired by the data sets we extracted the production trace patterns that can be used to enrich data sets with hidden information. The patterns are grouped by the type of information they retrieve. We distinguish the following types of patterns to enrich the production trace: `aggregation over an interval' (Section~\ref{sec:aggregate-over-interval}), `calculating the time that elapsed between events' (Section~\ref{sec:compute-time-delta}), `deriving relation between entity and event' (Section~\ref{sec:derive-relation-events}), and `deriving relation between entities' (Section~\ref{sec:derive-relation-entities}). Table \ref{tab:overview_patterns} gives an overview of the patterns, in which section they are discussed, what type of operation is used, and which of the classes from the model are represented.

\begin{table}[ht]
    \centering
    \renewcommand\arraystretch{1.3}
    \begin{tabular}{l|c|c|c|c|c}
        \multirow{2}{*}{\textbf{Section}} & \multirow{2}{*}{\textbf{Pattern}} & \multirow{2}{*}{\textbf{Operation}} & \multicolumn{3}{c}{\textbf{Class}} \\ \cline{4-6}
         & & & \texttt{Event} & \texttt{Resource} & \texttt{ProductionEntity} \\ \hline \hline
        \multirow{2}{*}{\ref{sec:aggregate-over-interval}} & \ref{pattern:interval_count} & Count & x & x & x \\ \cline{2-6}
         & \ref{pattern:interval_aggregate} & Aggregate & x & x & x \\ \hline
        \multirow{3}{*}{\ref{sec:compute-time-delta}} & \ref{pattern:elapsed-time_preceding} & Subtract & x & x & \\ \cline{2-6}
         & \ref{pattern:elapsed-time_same-type} & Subtract & x & x &  \\ \cline{2-6}
         & \ref{pattern:elapsed-time_maximum} & Subtract & x & x & x \\ \hline
        \multirow{4}{*}{\ref{sec:derive-relation-events}} & \ref{pattern:event-entity_preceding} & Relate & x & x &  \\ \cline{2-6}
         & \ref{pattern:event-entity_partOf} & Relate & x & x & x \\ \cline{2-6}
         & \ref{pattern:event-entity_all-preceding} & Relate & x &  & x \\ \cline{2-6}
         & \ref{pattern:event-entity_all-succeeding} & Relate & x &  & x \\ \hline
        \ref{sec:derive-relation-entities} & \ref{pattern:entity-entity_partOf} & Relate & x & x & x \\
    \end{tabular}
    \caption{Overview of the production trace patterns}
    \label{tab:overview_patterns}
\end{table}

In line with Suriadi et al. \cite{Suriadi2017EventLogs}, the following components are used to describe each pattern:
\begin{itemize}
    \item \textit{Definition}: each pattern at least includes the following elements:
    \subitem - An operation, like aggregate values or relate objects;
    \subitem - One or two target object types, in terms of the reference model;
    \subitem - A condition or constraint to target the relevant objects, for example a timing relation between events or the relation between an entity and an event.
    \item \textit{Description}: a textual description of the pattern and in what type of manufacturing settings it is typically observed.
    \item \textit{Use cases}: a set of use cases obtained from the different data sets (Table~\ref{tab:overview_data_sets}). Next to a description of the use case, the application of the pattern is illustrated through:
    \subitem - A semi-formal rule or instantiation of the pattern for each of the listed use cases, using the terminology as defined in the reference model.
    \subitem - A visualisation of the pattern applied to one of the provided use cases. We also describe the parameter values that can be used to instantiate the pattern for the visualized use case.
    \item \textit{Lightweight rule}: 'Lightweight' formal specification using SPARQL (query templates), which are shared in a Github repository\footnote{\url{https://github.com/gitmpje/production-trace-patterns/tree/main/sparql}}. Next to the formalised patterns, the repository also contains a dummy data set and a script to apply the patterns on the data set for some of the provided use cases.
\end{itemize}
The patterns that are presented in this work are derived from experiences analysing data sets (Section \ref{sec:patterns_data_sets}).

\subsection{Enrich trace by aggregation over an interval}
\label{sec:aggregate-over-interval}
The following patterns occur frequently in production environments with various resources executing different steps in the production process. In those environments, data are typically captured at different levels. On the one hand, there are many events that mark the start and end of an interval. On the other hand, there are various events taking place during that interval. Aggregating the events in the interval will reveal new information. For example, the events describing the track in and out (start and completion of manufacturing) of a job at a machine are typically captured by the MES, while events like machine alarms are captured by a different system, e.g. an alarm monitoring system. For analysis of the performance of the machine and root cause analysis of incidents, a useful insight is the number of machine alarms that took place during the processing of the job. Similarly, there are often sensors monitoring the machine, but data from a sensor often provide valuable insights only after aggregating them over a certain time window. Below we describe the patterns that can be used to derive different types of information over an interval.

\begin{pattern}
    Count \texttt{Event}s of a given type in an interval marked by two \texttt{Event}s of given types, related to the same \texttt{Resource} and \texttt{ProductionEntity}
    \label{pattern:interval_count}
\end{pattern}
\paragraph{Description} This pattern captures all use cases where new information can be derived by counting the events of a certain type in the interval between the start and finish of processing a production entity (like a job or batch) by a resource, such as the number of alarms that occurred while a machine was processing a job.\\

\paragraph{Use cases}
\begin{enumerate}[label=\thepattern-\arabic*]
    \item In the interval between the start and finish of processing a job by a machine: count all alarms.\label{uc:interval_alarms}\\
    \textit{Rule}: Count the number of \texttt{Alarm}s between a \texttt{TrackIn} and \texttt{TrackOut} for a \texttt{Job} on a \texttt{Machine}.
    \item In the interval between the start and finish of processing a production lot by a machine: count all machine repairs.\label{uc:interval_repairs}\\
    \textit{Rule}: Count the number of \texttt{Repair}s between a \texttt{TrackIn} and \texttt{TrackOut} for a \texttt{Job} on a \texttt{Machine}.
\end{enumerate}

\begin{figure}[H]
    \centering
    \includegraphics[width=0.75\linewidth]{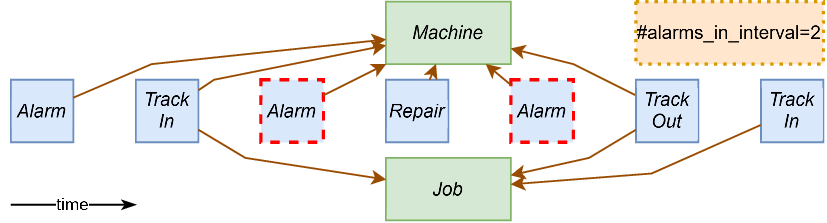}
    \caption{Illustration of Pattern~\ref{pattern:interval_count} applied to a data set containing alarms.}
    \label{fig:uc_interval_alarms}
\end{figure}

Figure~\ref{fig:uc_interval_alarms} gives an example of use case~\ref{uc:interval_alarms}, in this example two alarms occur in the interval between track in and out of a job on a machine. The (SPARQL) pattern template can be instantiated for this example as follows:
\begin{itemize}
    \item Interval start event type: \textit{IntervalStartType} = \texttt{TrackIn};
    \item Interval end event type: \textit{IntervalEndType} = \texttt{TrackOut};
    \item Type of event that should be counted: \textit{EventType} = \texttt{Alarm}.
\end{itemize}

\begin{pattern}
    Aggregate \texttt{Event} attribute in an interval marked by two \texttt{Event}s of given types, related to the same \texttt{Resource} and \texttt{ProductionEntity}
    \label{pattern:interval_aggregate}
\end{pattern}
\paragraph{Description} This pattern captures all use cases where new information can be derived by aggregating an attribute of the event in the interval between the start and finish of processing a production entity by a resource. In contrast to Pattern~\ref{pattern:interval_count}, this pattern aims to derive information by looking at specific attributes of the events that occur in the interval. A common example is the aggregation of sensor measurements, for example to detect if a certain parameter was on average higher when processing a job.\\

\begin{figure}[H]
    \centering
    \includegraphics[width=0.75\linewidth]{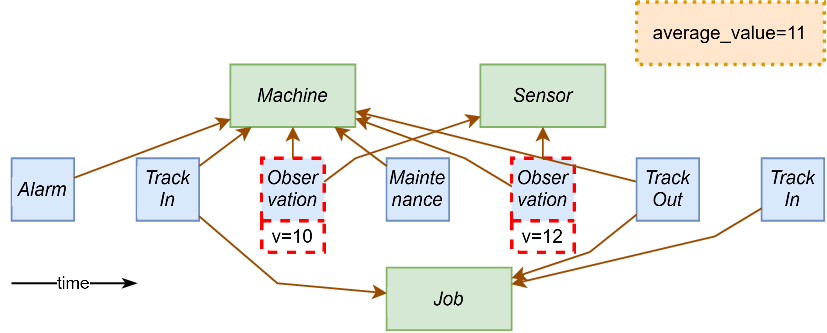}
    \caption{Example of the application of Pattern~\ref{pattern:interval_aggregate} on a data set with sensor observations.}
    \label{fig:uc_average-measurement}
\end{figure}

\paragraph{Use cases}
\begin{enumerate}[label=\thepattern-\arabic*]
    \item Calculate average sensor measurement while processing a job (Figure~\ref{fig:uc_average-measurement}). \label{uc:interval_average-measurement}\\
    \textit{Rule}: Calculate the average measured value (\texttt{v}) of all \texttt{Observation}s between a \texttt{TrackIn} and \texttt{TrackOut} for a \texttt{Job} on a \texttt{Machine}.
    \item Count the number of times a sensor value crosses a certain threshold while processing a job.\label{uc:interval_cross-threshold}\\
    \textit{Rule}: Count the number of times the value of an \texttt{Observation} between a \texttt{TrackIn} and \texttt{TrackOut} for a \texttt{Job} on a \texttt{Machine} is above a given threshold.
    \item Sum the number of entities that are rejected between the start and finish of processing a production lot.\label{uc:interval_rejects}\\
    \textit{Rule}: Sum the \texttt{quantityRejected} values between the \texttt{TrackIn} and \texttt{TrackOut} of that \texttt{ProductionLot} at a \texttt{Workstation}.
\end{enumerate}

Figure~\ref{fig:uc_average-measurement} gives an example of use case~\ref{uc:interval_average-measurement}, in this example two observations are made, with an average value of 11. The (SPARQL) pattern template can be instantiated for this example as follows:
\begin{itemize}
    \item Interval start event type: \textit{IntervalStartType} = \texttt{TrackIn};
    \item Interval end event type: \textit{IntervalEndType} = \texttt{TrackOut};
    \item Event type of interest: \textit{EventType} = \texttt{Observation};
    \item Attribute of event that should be aggregated: \textit{attribute} = \texttt{value} (\texttt{v}).
\end{itemize}

\subsection{Enrich trace by calculating the time that elapsed between events}
\label{sec:compute-time-delta}
Next to defining an interval based on two events, the time that elapsed between certain events is useful information that can be derived from manufacturing event data sets. An obvious use case is to calculate the time between the start and finish of processing a job by a resource, which gives information about the performance of the resource and can be compared with the expected or planned processing time. Likewise, for the time between the start and end time of a machine breakdown. In other cases, it could be interesting to know the time between when a machine started processing a job and the closest preceding repair/maintenance activity, which might have impacted the machine's performance (in terms of throughput and quality).

\begin{pattern}
    Calculate the time between an \texttt{Event} and the closest preceding \texttt{Event} of given types related to a \texttt{Resource}
    \label{pattern:elapsed-time_preceding}
\end{pattern}
\paragraph{Description} This pattern captures all use cases where new information can be derived by comparing an event and the closest preceding event for a resource.\\

\paragraph{Use cases}
\begin{enumerate}[label=\thepattern-\arabic*]
    \item Calculate the time (\texttt{t}) it took to process a job on a machine.\label{uc:compute-time_processing}\\
    \textit{Rule}: Calculate the time between a \texttt{TrackOut} and the closest preceding \texttt{TrackIn} for a \texttt{Job} on a \texttt{Machine}.
    \item Calculate the time since the last maintenance before start processing a job.\label{uc:compute-time_maintenance}\\
    \textit{Rule}: Calculate the time between a \texttt{TrackIn} for a \texttt{Job} on a \texttt{Resource} and the closest preceding \texttt{Maintenance} for that \texttt{Resource}.
    \item Calculate the time between processing two jobs (setup time).\label{uc:compute-time_setuptime}\\
    \textit{Rule}: Calculate the time between a \texttt{TrackOut} and the closest preceding \texttt{TrackIn} for a \texttt{Resource}.
    \item Calculate the buffer time of a product in front of a resource.\label{uc:compute-time_buffer}\\
    \textit{Rule}: Calculate the time between a \texttt{TrackOut} and the closest preceding \texttt{TrackIn} for a \texttt{Product} on a \texttt{Buffer}.
    \item Calculate the transportation time of a product by an AGV.\label{uc:compute-time_transport}\\
    \textit{Rule}: Calculate the time between a \texttt{TrackOut} and the closest preceding \texttt{TrackIn} for a \texttt{Product} on a \texttt{AGV}.
\end{enumerate}

\begin{figure}[H]
    \centering
    \includegraphics[width=0.75\linewidth]{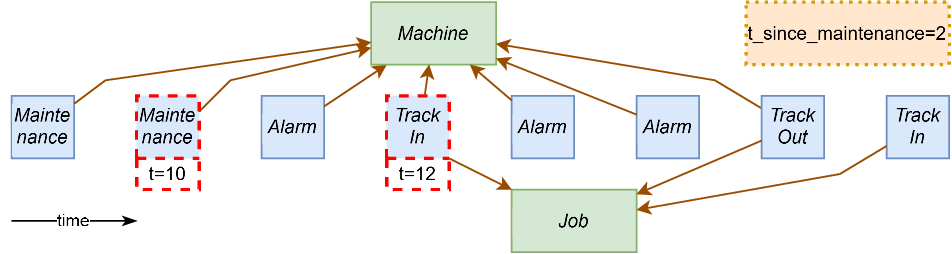}
    \caption{Illustration of applying Pattern~\ref{pattern:elapsed-time_preceding} to a data set containing maintenance events.}
    \label{fig:uc_compute-time_maintenance}
\end{figure}

Figure~\ref{fig:uc_compute-time_maintenance} gives an example of use case~\ref{uc:compute-time_maintenance}, in this example the time since the last maintenance can be derived from the track-in event with timestamp 12 and the maintenance event with timestamp 10. The (SPARQL) pattern template can be instantiated for this example as follows:
\begin{itemize}
    \item Event type of interest: \textit{EventType} = \texttt{TrackIn};
    \item Preceding event type of interest: \textit{PrecedingEventType} = \texttt{Maintenance}.
\end{itemize}

\begin{pattern}
    Calculate the time between an \texttt{Event} and the closest succeeding \texttt{Event} of the same type related to a \texttt{Resource}
    \label{pattern:elapsed-time_same-type}
\end{pattern}
\paragraph{Description} This pattern captures all use cases where new information can be derived by calculating the time that elapsed between two subsequent events of the same type. An example application is the calculation of the downtime of a machine, which can be derived from the time between when the machine switched to a failed state and the following event where it switches back to another state.\\

\begin{figure}[H]
    \centering
    \includegraphics[width=0.75\linewidth]{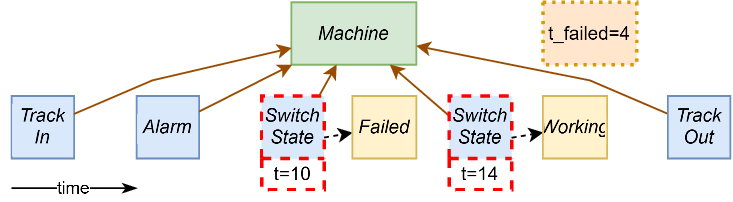}
    \caption{Illustration of applying Pattern~\ref{pattern:elapsed-time_same-type} to a data set containing switch state events.}
    \label{fig:uc_compute-time_downtime}
\end{figure}

\paragraph{Use cases}
\begin{enumerate}[label=\thepattern-\arabic*]
    \item Calculate the downtime of a machine (Figure~\ref{fig:uc_compute-time_downtime}).\label{uc:compute-time_downtime}\\
    \textit{Rule}: Calculate the time between a \texttt{SwitchState} to \texttt{Failed} state and the closest succeeding \texttt{SwitchState} for a \texttt{Machine}.
\end{enumerate}

Figure~\ref{fig:uc_compute-time_downtime} gives an example of use case~\ref{uc:compute-time_downtime}. In this example the downtime can be derived from the switch to failed state at time 10 and the switch back to working state at time 14. The (SPARQL) pattern template can be instantiated for this example as follows:
\begin{itemize}
    \item Event type of interest: \textit{EventType} = \texttt{SwitchState}.
\end{itemize}

\begin{pattern}
    Calculate the maximum time between two \texttt{Event}s of given types related to an \texttt{Entity}
    \label{pattern:elapsed-time_maximum}
\end{pattern}
\paragraph{Description} In addition to the time between two subsequent events, it can also be of interest to look at the maximum time between two events of a certain type. This pattern can for example be applied to calculate the throughput time of a product, which can be derived from the first and last event where this product was tracked at a resource.

\paragraph{Use cases}
\begin{enumerate}[label=\thepattern-\arabic*]
    \item Calculate the throughput time of a product.\label{uc:compute-time_throughput}\\
    \textit{Rule}: Calculate the maximum time between a \texttt{TrackIn} and \texttt{TrackOut} of a \texttt{Product}.
\end{enumerate}

\begin{figure}[H]
    \centering
    \includegraphics[width=0.75\linewidth]{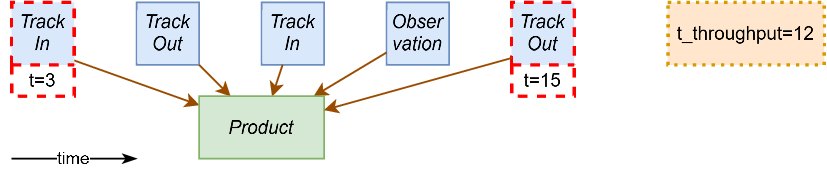}
    \caption{Illustration of applying Pattern~\ref{pattern:elapsed-time_maximum} to a series of events for a product.}
    \label{fig:uc_compute-time_throughput}
\end{figure}

Figure~\ref{fig:uc_compute-time_throughput} gives an example of use case~\ref{uc:compute-time_throughput}. In this example the throughput time can be derived from the first track in at time 3 and the last track out at time 15. The (SPARQL) pattern template can be instantiated for this example as follows:
\begin{itemize}
    \item Interval start event type: \textit{IntervalStartType} = \texttt{TrackIn};
    \item Interval end event type: \textit{IntervalEndType} = \texttt{TrackOut}.
\end{itemize}

\subsection{Enrich trace by deriving a relation between an event and entity}
\label{sec:derive-relation-events}
When data from different sources are combined there are often hidden relations that can be derived. One typical use case is that it is interesting to know what tool is used by a resource to process a job. This relation can be derived by combining the switch tool events for the resource with the events describing the processing of a job by that resource. In other cases, it is possible to derive information about the higher aggregation level entities, like a batch, from events related to lower level entities, like a product, when it is known that this product was part of the batch. In that case, the relation between the lower-level events and the higher-level entity can be derived. This information can be used to generate a more complete view of the production process for a batch, and can for example be used to calculate the total processing time of the batch.

\begin{pattern}
    Relate \texttt{Event} ($e1$) to a \texttt{Resource} that is related to the closest preceding \texttt{Event} ($e2$) of a given type
    \label{pattern:event-entity_preceding}
\end{pattern}
\paragraph{Description} This pattern captures all use cases where new information can be derived by finding an event closest preceding the start processing at a resource and its related entities.\\

\paragraph{Use cases}
\begin{enumerate}[label=\thepattern-\arabic*]
    \item Derive the relation between the event that marks the start of processing a job on a machine and the tool related to the closest preceding switch tool event.\label{uc:event-entity_tool}\\
    \textit{Rule}: Derive the relation between a \texttt{TrackIn} of a \texttt{Job} on a \texttt{Machine} and the \texttt{Tool} related to the closest preceding \texttt{SwitchTool} for that \texttt{Machine}.
\end{enumerate}

\begin{figure}[H]
    \centering
    \includegraphics[width=0.75\linewidth]{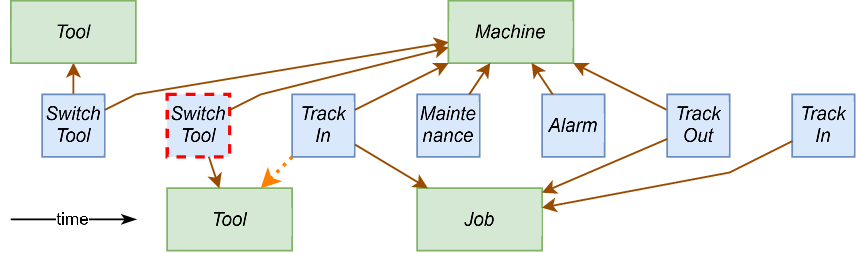}
    \caption{Illustration of applying Pattern~\ref{pattern:event-entity_preceding} to a data set containing switch tool events.}
    \label{fig:uc_event-entity_tool}
\end{figure}

Figure~\ref{fig:uc_event-entity_tool} gives an example of use case~\ref{uc:event-entity_tool}, where the machine switched tools directly before a job was tracked at that machine. The (SPARQL) pattern template can be instantiated for this example as follows:
\begin{itemize}
    \item Event type of interest: \textit{EventType} = \texttt{TrackIn};
    \item Preceding event type of interest: \textit{PrecedingEventType} = \texttt{SwitchTool}.
\end{itemize}

\begin{pattern}
    Relate \texttt{Event} to an \texttt{Entity} based on a \texttt{isPartOf} relation between \texttt{Entity}s
    \label{pattern:event-entity_partOf}
\end{pattern}
\paragraph{Description} This pattern captures all use cases where new information can be derived by finding all events related to a production entity that is part of another production entity. Discovering the relation between lower-level entities and production events is useful for traceability, it can for example be used to discover the production trace for a specific product, instead of the production lot this product is part of. Next to that, it can give insights into the performance of resources, for example the number (and type) of products that a resource worked on during a certain interval.\\

\paragraph{Use cases}
\begin{enumerate}[label=\thepattern-\arabic*]
    \item Derive the relation between an event on the production lot level and the product that is part of that lot.\label{uc:event-entity_product-lot}\\
    \textit{Rule}: Derive the relation between an \texttt{Event} related to a \texttt{ProductionLot} and the \texttt{Product} which \texttt{isPartOf} that \texttt{ProductionLot}.
    \item Derive the relation between an observation made by a sensor and the resource where this sensor is located.\label{uc:event-entity_sensor-resource}\\
    \textit{Rule}: Derive the relation between an \texttt{Observation} by a \texttt{Sensor} and the \texttt{Resource} that this \texttt{Sensor} \texttt{isPartOf}.
\end{enumerate}

\begin{figure}[H]
    \centering
    \includegraphics[width=0.5\linewidth]{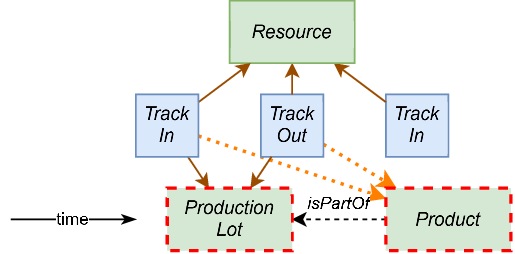}
    \caption{Application of Pattern~\ref{pattern:event-entity_partOf} on product and production lot level events.}
    \label{fig:uc_event-entity_product-lot}
\end{figure}

Figure~\ref{fig:uc_event-entity_product-lot} gives an example of use case~\ref{uc:event-entity_product-lot}, where there are events captured on different aggregation levels. The (SPARQL) pattern template can be instantiated without specifying parameters.

It is common in batch/lot manufacturing environments that lots are split into smaller lots and/or merged into bigger lots, for example to distribute the load over the different machines optimally \cite{Cheng2013AStreaming}. The meaning of a split or merge event can be used to relate an event to other entities it is (possibly) related to. For example, if some entities are split from a lot into a new lot, then the events related to this new lot are also related to (entities that were part of) the initial lot. Note that this can be applied recursively, to create the relevant relations along the complete chain of aggregation (split/merge/consume) events. This use case generally applies to aggregation events \cite{GS12022EPCISStandard}.

\begin{pattern}
    Relate \texttt{Event}s preceding an \texttt{Aggregate} event to an indirectly related \texttt{ProductionEntity}
    \label{pattern:event-entity_all-preceding}
\end{pattern}
\paragraph{Description} This pattern captures all use cases where new information can be derived by finding all events preceding an event where some production entities are aggregated. For example, consider two production lots that first follow distinct production traces on different machines, are subsequently merged into one production lot, and then continue as one entity through the remaining production steps. In this scenario, Pattern~\ref{pattern:event-entity_all-preceding} can be applied to relate all events from the distinct production lots to the merged production lot and in that manner discover the complete production trace for the products in the merged production lot.\\

\paragraph{Use cases}
\begin{enumerate}[label=\thepattern-\arabic*]
    \item Derive the relation between events related to a lot, and the lots split from that lot.\label{uc:event-entity_original-split}\\
    \textit{Rule}: Derive the relation between \texttt{Event}s preceding a \texttt{Split} related to a \texttt{ProductionLot} and another \texttt{ProductionLot} that is also related to this \texttt{Split}.
    \item Derive the relation between events related to a lot that was merged into another lot, and the merged lot.\label{uc:event-entity_original-merged}\\
    \textit{Rule}: Derive the relation between \texttt{Event}s preceding a \texttt{Merge} related to a \texttt{ProductionLot} and another \texttt{ProductionLot} that is also related to this \texttt{Merge}.
    \item Derive the relation between events related to a component that is consumed to assemble/produce to produce a product.\label{uc:event-entity_material-product}\\
    \textit{Rule}: Derive the relation between \texttt{Event}s preceding a \texttt{Consume} related to a \texttt{Component} and a \texttt{Product} that is also related to this \texttt{Consume}.
\end{enumerate}

\begin{figure}[H]
    \centering
    \includegraphics[width=0.75\linewidth]{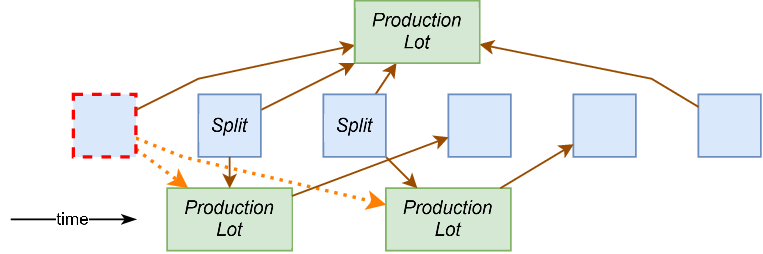}
    \caption{Illustration of applying Pattern~\ref{pattern:event-entity_all-preceding} to a data set containing an event describing the split of a production lot into smaller lots.}
    \label{fig:uc_event-entity_original-split}
\end{figure}

Figure~\ref{fig:uc_event-entity_original-split} gives an example of use case~\ref{uc:event-entity_original-split}, where there is one event related to the lot before two lots are split from that lot. The (SPARQL) pattern template can be instantiated for this example as follows:
\begin{itemize}
    \item Type of entity for which new relations should be derived: \textit{RelatedEntityType} = \texttt{ProductionLot}.
\end{itemize}

\begin{pattern}
    Relate \texttt{Event}s succeeding an \texttt{Aggregate} event to an indirectly related \texttt{ProductionEntity}
    \label{pattern:event-entity_all-succeeding}
\end{pattern}
\paragraph{Description} This pattern captures all use cases where new information can be derived by finding all succeeding events related to a production entity that is involved in the same event as another production entity.\\

\paragraph{Use cases}
\begin{enumerate}[label=\thepattern-\arabic*]
    \item Derive the relation between events related to a lot that was split from a lot and the original lot.\label{uc:event-entity_split-original}\\
    \textit{Rule}: Derive the relation between \texttt{Event}s succeeding a \texttt{Split} related to a \texttt{ProductionLot} and another \texttt{ProductionLot} that is also related to this \texttt{Split}.
    \item Derive the relation between events related to a merged lot, and the lots merged into that lot.\label{uc:event-entity_merged-original}\\
    \textit{Rule}: Derive the relation between \texttt{Event}s succeeding a \texttt{Merge} related to a \texttt{ProductionLot} and another \texttt{ProductionLot} that is also related to this \texttt{Merge}.
\end{enumerate}

\begin{figure}[H]
    \centering
    \includegraphics[width=0.75\linewidth]{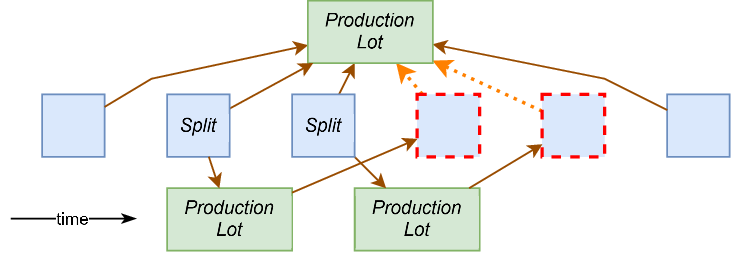}
    \caption{Illustration of applying Pattern~\ref{pattern:event-entity_all-succeeding} to a data set containing an event describing the split of a production lot into smaller lots.}
    \label{fig:uc_event-entity_split-original}
\end{figure}

Figure~\ref{fig:uc_event-entity_split-original} gives an example of use case~\ref{uc:event-entity_split-original}, where there are two events related to lots that were split from another lot. The (SPARQL) pattern template can be instantiated for this example as follows:
\begin{itemize}
    \item Type of entity for which new relations should be derived: \textit{RelatedEntityType} = \texttt{ProductionLot}.
\end{itemize}

\subsection{Enrich trace by deriving a relation between two entities}
\label{sec:derive-relation-entities}
In an environment where data from various aggregation levels are collected by isolated systems, it might remain unknown whether a specific product is part of a batch. For example, the batch-level events will be captured by the MES, while there might be another system that registers operations on individual products by a machine. These data can be combined to derive to what batch the different products belong.

\begin{pattern}
    Relate \texttt{ProductionEntities} (\texttt{partOf} relation) based on \texttt{Event}s between \texttt{TrackIn} and \texttt{TrackOut} at a \texttt{Resource}
    \label{pattern:entity-entity_partOf}
\end{pattern}
\paragraph{Description} This pattern captures all use cases where new information can be derived by finding all production entities that are part of another production entity based on events that occurred in a certain interval. Deriving the 'part of' relation is in the first place useful for traceability, where it is necessary to know which product belonged to which batch(es). The relation can also be used for monitoring the performance, in which case we are interested in the number of products that are part of a batch.\\

\paragraph{Use cases}
\begin{enumerate}[label=\thepattern-\arabic*]
    \item Derive the logical 'part of' relation between a product and a batch from (lower level) events related to the product in the interval between the start and finish of processing the batch on a resource.\label{uc:entity-entity_batch-products}\\
    \textit{Rule}: Derive the \texttt{isPartOf} relation between a \texttt{Product} and a \texttt{Batch} from an \texttt{Event} related to the \texttt{Product} that occurs in the interval between \texttt{TrackIn} and \texttt{TrackOut} of the \texttt{Batch} at a \texttt{Resource}.
    \item Derive the physical 'part of' relation between a component and a product from consume/assembly events in the interval between the start and finish of processing the product.\label{uc:entity-entity_product-material}\\
    \textit{Rule}: Derive the \texttt{isPartOf} relation between a \texttt{Product} and a \texttt{Component} from a \texttt{Consume} related to the \texttt{Component} that occurs in the interval between \texttt{TrackIn} and \texttt{TrackOut} of the \texttt{Product} at a \texttt{Resource}.
\end{enumerate}

\begin{figure}[H]
    \centering
    \includegraphics[width=0.5\linewidth]{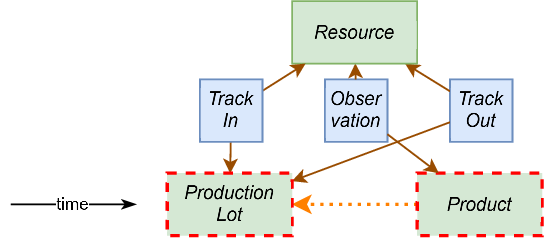}
    \caption{Application of Pattern~\ref{pattern:entity-entity_partOf} on product and production lot level events.}
    \label{fig:uc_entity-entity_batch-products}
\end{figure}

Figure~\ref{fig:uc_entity-entity_batch-products} gives an example of use case~\ref{uc:entity-entity_batch-products}, where a specific product is observed at a resource in the interval where this resource was processing a certain production lot. In this case it can be derived that the product is part of the production lot. The (SPARQL) pattern template can be instantiated for this example as follows:
\begin{itemize}
    \item Interval start event type: \textit{IntervalStartType} = \texttt{TrackIn};
    \item Interval end event type: \textit{IntervalEndType} = \texttt{TrackOut};
    \item Type of entity (on a lower aggregation level) for which to derive the part of relation: \textit{PartEntityType} = \texttt{Product}.
\end{itemize}

\subsection{Combining patterns}
The patterns listed above can also be combined and sequentially applied to derive further (aggregated) information. Below we describe two example use cases and what patterns can be used to derive the information.

\textbf{Average downtime of a resource}: a typical piece of information that is of interest for monitoring the performance of resources is the average downtime of the resource. Use case~\ref{uc:compute-time_downtime} already showed how Pattern~\ref{pattern:elapsed-time_same-type} can be used to compute the duration of one downtime period of a machine. Subsequently, Pattern~\ref{pattern:interval_aggregate} can be used to aggregate the downtime of the machine over a certain interval, for example the average or variation.

\textbf{Aggregated average processing time}: similarly, Pattern~\ref{pattern:interval_aggregate} can be combined with Pattern~\ref{pattern:elapsed-time_maximum} to compute the average processing time of a resource, which is another useful insight for monitoring the performance.

\section{Evaluation}
\label{sec:evaluation}
In this section, we present empirical evidence supporting the claims that the patterns are valuable for knowledge exchange and for accelerating data aggregation across a range of manufacturing environments. To this end, we demonstrate the applicability of the proposed patterns in diverse datasets and by conducting semi-structured interviews with professionals from multiple industrial sectors.

\subsection{Applicability}
First, we evaluated the applicability of the patterns by recording their appearance in the data sets. We consider the use cases presented in the previous section and the manufacturing data sets presented in Section~\ref{sec:patterns_data_sets}. The results of the study are summarised in Table~\ref{tab:overview_examples_use-cases}, which gives an overview of the use cases for the patterns and the data sets in which they are observed. As can be seen in Table~\ref{tab:overview_examples_use-cases}, most of the use cases are observed in multiple data sets. If we aggregate the results over the patterns, we can see that most of the patterns apply to at least three data sets, except for Pattern~\ref{pattern:elapsed-time_same-type}, \ref{pattern:event-entity_preceding}, and~\ref{pattern:event-entity_all-succeeding}.

A use case for Pattern~\ref{pattern:elapsed-time_same-type} is the downtime of production resources, which is also relevant in other settings but is captured by different types of events. In these settings, Pattern~\ref{pattern:elapsed-time_preceding} can be used to derive the information. Pattern~\ref{pattern:event-entity_preceding} and \ref{pattern:event-entity_all-succeeding} are more specific for production processes in which large volumes of products are processed in batches, which we only observed in one of the data sets.

To illustrate the practical use of the proposed patterns, we implemented them as SPARQL \cite{TheW3CSPARQLWorkingGroup2013SPARQLOverview} query templates\footnote{\url{https://github.com/gitmpje/production-trace-patterns}}. In addition, we developed a script that automatically instantiates and executes these templates on the anonymised automotive data set, thereby enabling a systematic and reproducible application of the patterns to real-world data.

\begin{table}[H]
\centering
\caption{Overview of the use cases and in what manufacturing setting they are observed.}
\label{tab:overview_examples_use-cases}
\begin{tabular}{c|c|c|c|c}
  \textbf{\begin{tabular}[c]{@{}c@{}}Use\\ case\end{tabular}} & \textbf{\begin{tabular}[c]{@{}c@{}}Semi-conductor\\ back-end\end{tabular}} & \textbf{Automotive} & \textbf{\begin{tabular}[c]{@{}c@{}}Contract\\ manufacturing\end{tabular}} & \textbf{\begin{tabular}[c]{@{}c@{}}Industrial\\ equipment\end{tabular}} \\ \hline \hline
  \rowcolor{lightgray}\ref{uc:interval_alarms} & x &  &  &  \\ \hline
  \rowcolor{lightgray}\ref{uc:interval_repairs} & x & x &  & x \\ \hline
  \ref{uc:interval_average-measurement} & x & x &  & x \\ \hline
  \ref{uc:interval_cross-threshold} & x &  &  & x \\ \hline
  \ref{uc:interval_rejects} & x &  &  &  \\ \hline
  \rowcolor{lightgray}\ref{uc:compute-time_processing} & x & x & x & x \\ \hline
  \rowcolor{lightgray}\ref{uc:compute-time_maintenance} & x & x & x &  \\ \hline
  \rowcolor{lightgray}\ref{uc:compute-time_setuptime} &  & x &  &  \\ \hline
  \rowcolor{lightgray}\ref{uc:compute-time_buffer} &  & x &  & x \\ \hline
  \ref{uc:compute-time_downtime} &  & x &  &  \\ \hline
  \rowcolor{lightgray}\ref{uc:compute-time_throughput} & x & x & x & x \\ \hline
  \ref{uc:event-entity_tool} &  & x &  &  \\ \hline
  \rowcolor{lightgray}\ref{uc:event-entity_product-lot} & x &  &  & x \\ \hline
  \rowcolor{lightgray}\ref{uc:event-entity_sensor-resource} & x & x &  & \\ \hline
  \ref{uc:event-entity_original-split} & x &  &  &  \\ \hline
  \ref{uc:event-entity_original-merged} & x &  &  &  \\ \hline
  \ref{uc:event-entity_material-product} & x & x &  & x \\ \hline
  \rowcolor{lightgray}\ref{uc:event-entity_split-original} & x &  &  &  \\ \hline
  \rowcolor{lightgray}\ref{uc:event-entity_merged-original} & x &  &  &  \\ \hline
  \ref{uc:entity-entity_batch-products} & x &  &  &  \\ \hline
  \ref{uc:entity-entity_product-material} & x & x & x & x \\ \hline
\end{tabular}
\end{table}

\subsection{User acceptance}
An evaluation of the user acceptance of the reference model with the production trace patterns was conducted via semi-structured interviews followed by an explanation of the approach with two examples of the patterns. Subsequently, participants were asked to complete a self-guided questionnaire to evaluate user acceptance on the three dimensions of the Technology Acceptance Model (TAM) \cite{Davis1989TechnologyTAM}: Perceived Usefulness, Perceived Ease of Use, and Behavioural Intention to Use. The questionnaire consists of 14 statements for which participants had to provide a score on a Likert scale from 1 (`Strongly Disagree') to 5 (`Strongly Agree'). The complete list of questions for the semi-structured interview and questionnaire can be found in \ref{app:questions}.

\begin{figure}[p]
    \centering
    \includegraphics[width=\linewidth]{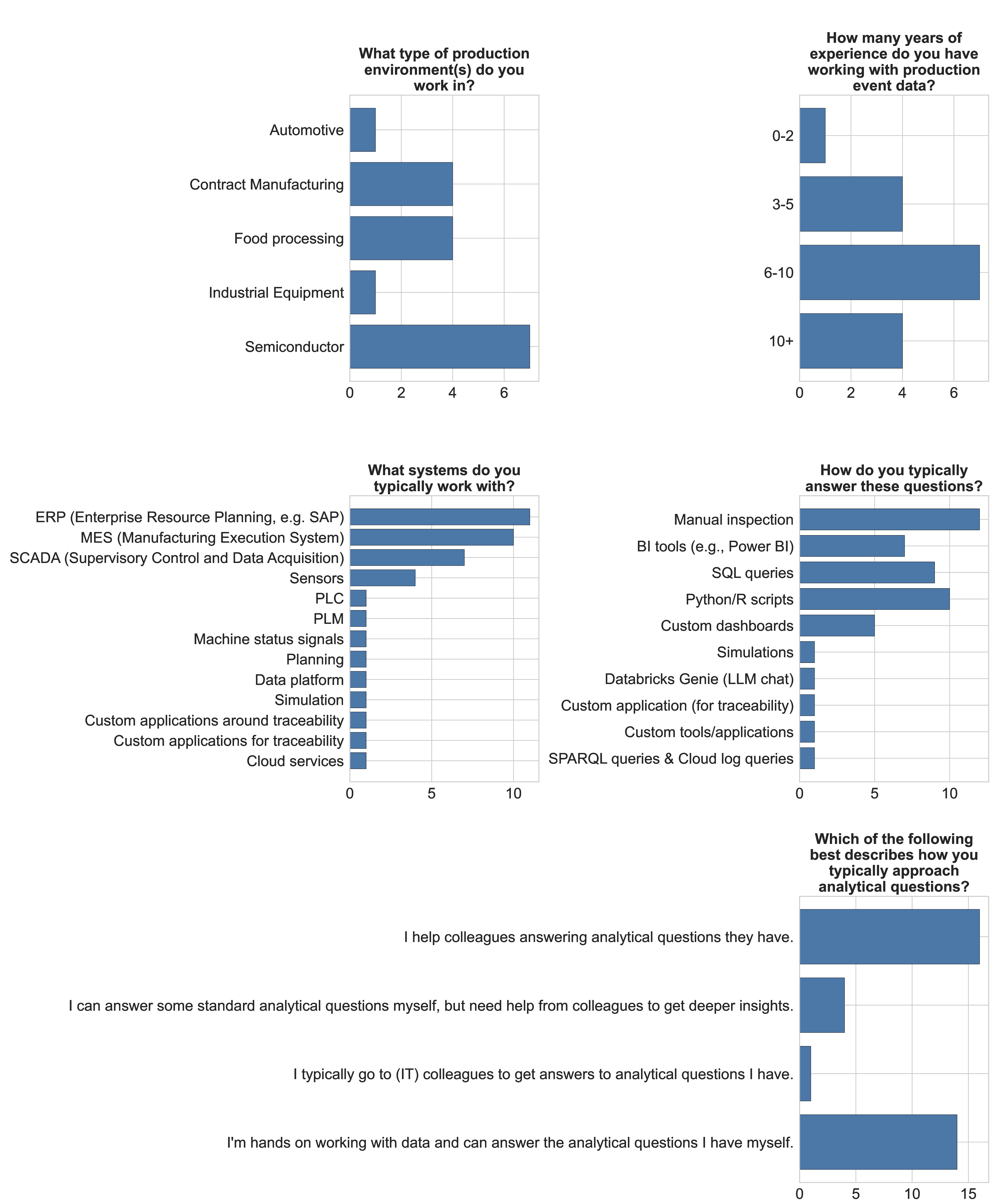}
    \caption{Overview of the answers to questions describing the experience of the participants.}
    \label{fig:barcharts_multiple_choice}
\end{figure}

The participants were selected from the researchers' network. In total, 16 professionals were interviewed, working in 13 different companies in the manufacturing domain. All participants work weekly to daily with production (event) data, and most participants have a mix of an IT- and business-oriented role. Figure~\ref{fig:barcharts_multiple_choice} shows the characteristics of the professionals. The participants work with a variety of tools in different manufacturing settings, but the majority works in the semiconductor industry. One participant has less than two years of experience, while four of the participants have over a decade of expertise. A complete overview of the descriptive information can be found in \ref{app:descriptives}.

\begin{figure}[H]
    \centering
    \includegraphics[width=0.7\textwidth]{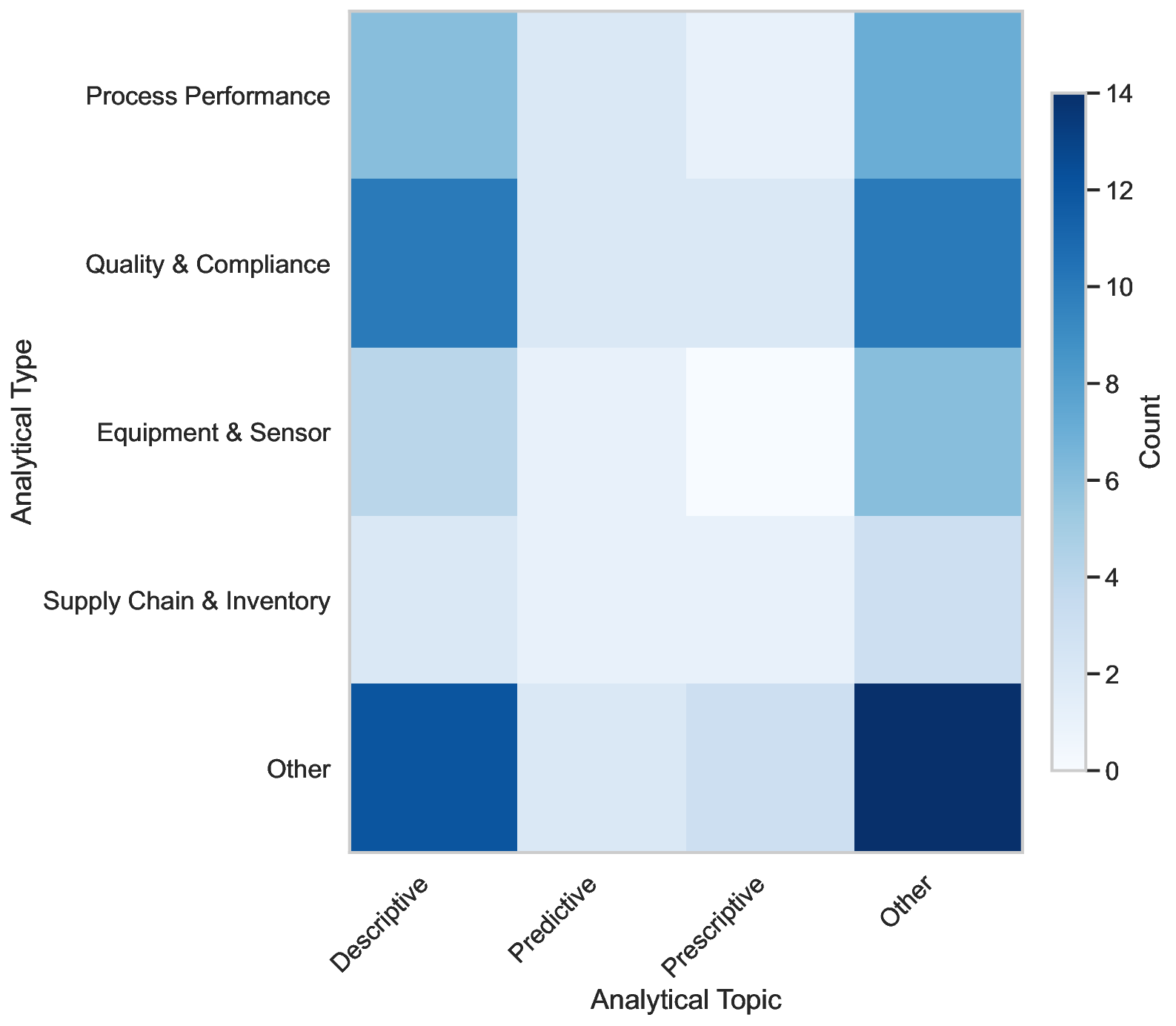}
    \caption{Co-occurrence of analytical topics and types in the answers to question 11 (`What analytical questions do you commonly ask?').}
    \label{fig:cooccurrence_themes_question11}
\end{figure}

We coded the answers to open question 11 on two dimensions: the analytical topic and the analytical type (descriptive, predictive, prescriptive). Figure~\ref{fig:cooccurrence_themes_question11} shows the co-occurrence of topics and types. Most common are descriptive analytical questions on the process performance (e.g. throughput time analysis) and quality \& compliance (e.g. statistical process control on faulty products). This is in line with the production trace patterns, which aim to accelerate and share descriptive insights. Questions about equipment \& sensor data (e.g. analysis of machine signals), and supply chain \& inventory (e.g. analysis of stocks in warehouses) are less common in the work of the participants. Because the responses vary so widely, many questions are categorised as `Other'. Separate counts for topics and analytical types can be found in \ref{app:themes_analytical_questions}.

\begin{figure}[H]
    \centering
    \begin{subfigure}[b]{0.48\textwidth}
        \centering
        \includegraphics[width=\textwidth]{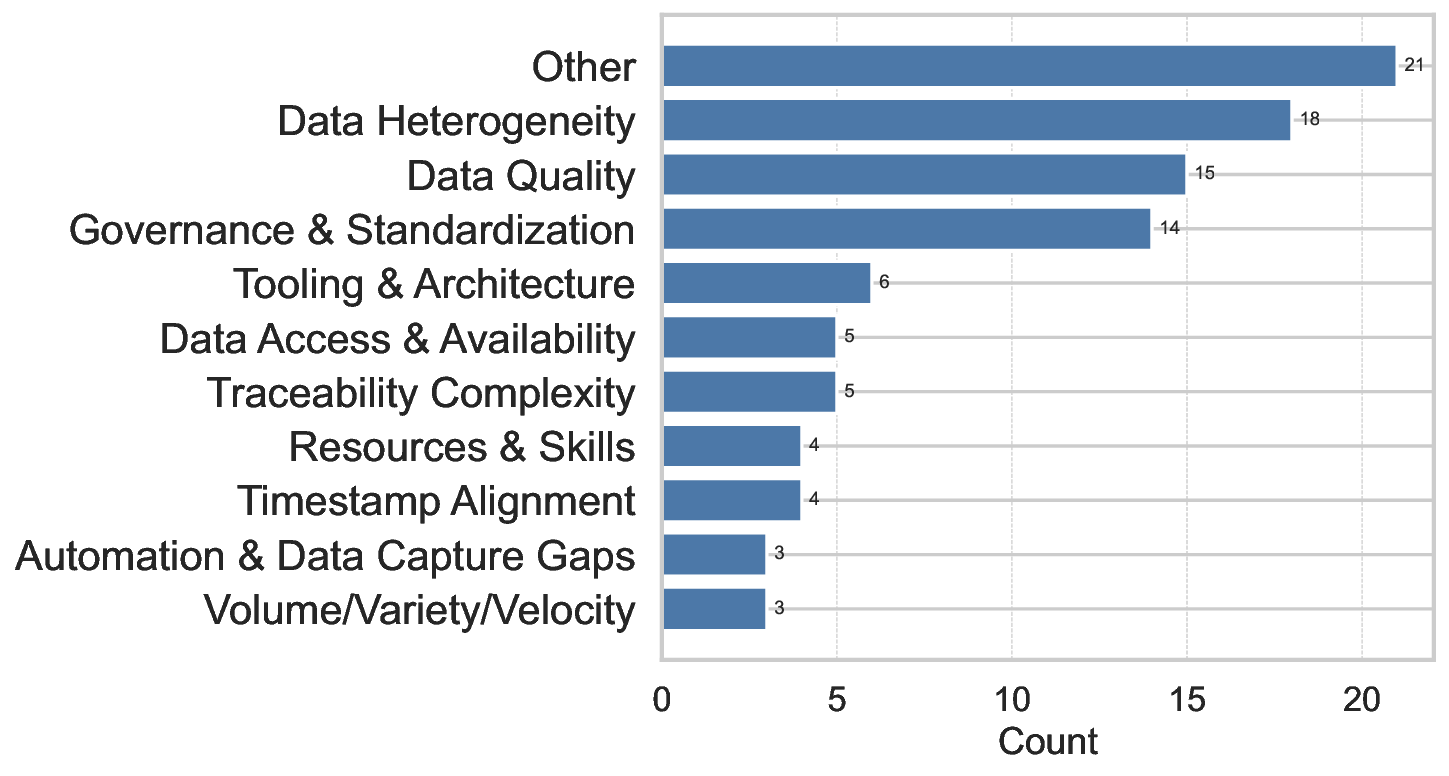}
        \caption{Occurrence of detailed themes.}
        \label{fig:detailed_themes_question12}
    \end{subfigure}
    \hfill
    \begin{subfigure}[b]{0.48\textwidth}
        \centering
        \includegraphics[width=\textwidth]{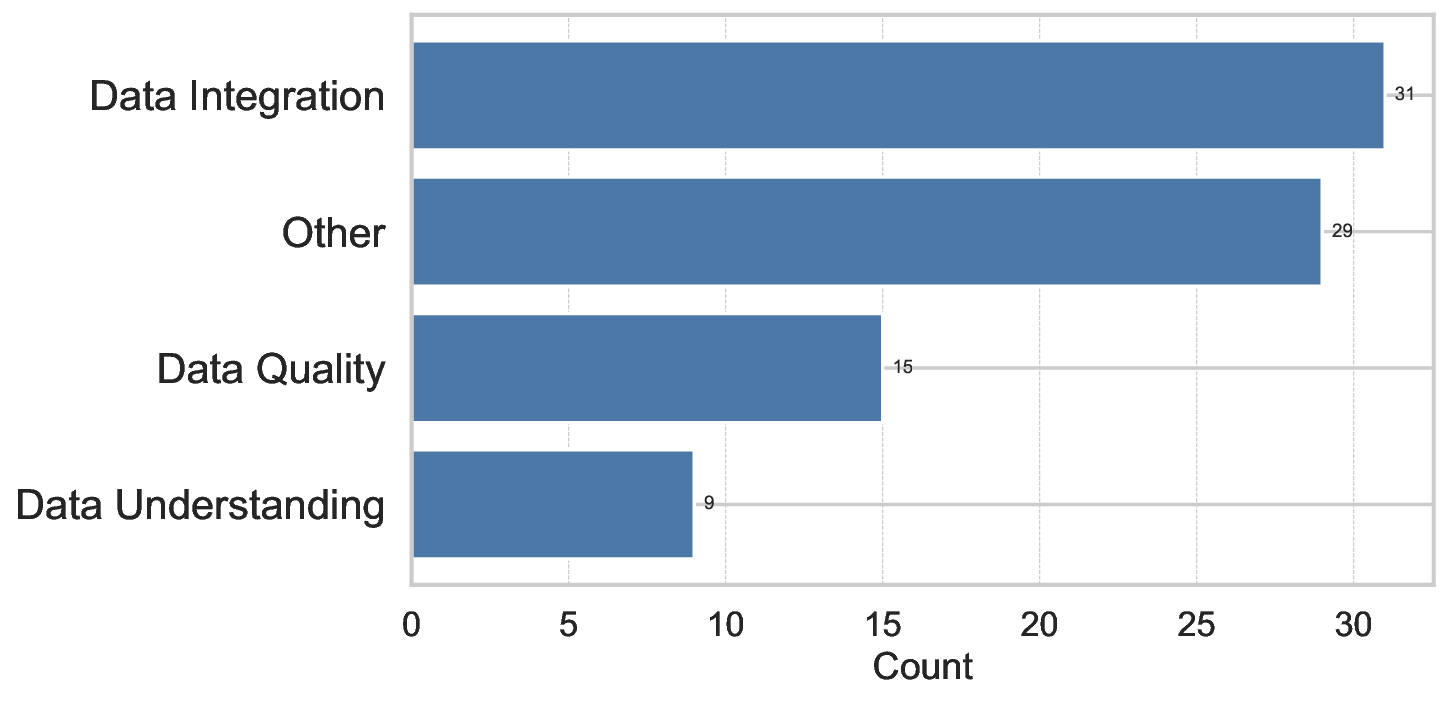}
        \caption{Occurrence of macro themes.}
        \label{fig:macro_themes_question12}
    \end{subfigure}
    
    \caption{Occurrence of themes in the answers to question 12 (`What are the biggest challenges you face when analyzing production event data? And/or what factors help to deal with those challenges?').}
    \label{fig:themes_question12}
\end{figure}

Table~\ref{tab:themes_question12} gives an overview of the themes on which we coded the answers to open question 12. Figures~\ref{fig:themes_question12} show the occurrence of themes in the responses of the participants. Common challenges perceived by the practitioners are: data understanding (e.g. lack of naming conventions), data quality (e.g. due to manual data entry), and data integration (e.g. inconsistencies in terms and identifiers used by different systems). The figure in \ref{app:cooccurrence_question_11_12} shows the frequency with which analytical topics and challenges co-occur.

\begin{table}[H]
    \centering
    \setlength{\tabcolsep}{3pt} 
    \begin{tabular}{p{0.28\textwidth}|p{0.24\textwidth}|p{0.40\textwidth}}
        \textbf{Challenge theme} & \textbf{Macro theme} & \textbf{Example} \\ \hline
        Data Quality & Data Quality & Missing/inconsistent data entries \\ \hline
        Data Heterogeneity & Data Integration & Different identifiers across systems and sites \\ \hline
        Traceability Complexity & Data Understanding & Complex data structures due to splits/merges in the process \\ \hline
        Data Governance \& Standardisation & Data Understanding & Documentation of data sources \\ \hline
        Data Access \& Availability & Other & Arranging access to data sources \\ \hline
        Volume/Variety/ Velocity & Data Integration & Handling big data sets \\ \hline
        Tooling \& Architecture & Data Integration & Different data storage formats \\ \hline
        Automation \& Data Capture Gaps & Other & Machine data not automatically stored \\ \hline
        Timestamp Alignment & Data Integration & Differences in timestamps used in different systems \\ \hline
        Resources \& Skills & Data Understanding & Data literacy
    \end{tabular}
    \caption{Overview of coding themes for the challenges reported under question 12 (``What are the biggest challenges you face when analyzing production event data? And/or what factors help to deal with those challenges?'')}
    \label{tab:themes_question12}
\end{table}

\begin{figure}[H]
    \centering
    \includegraphics[width=0.75\linewidth]{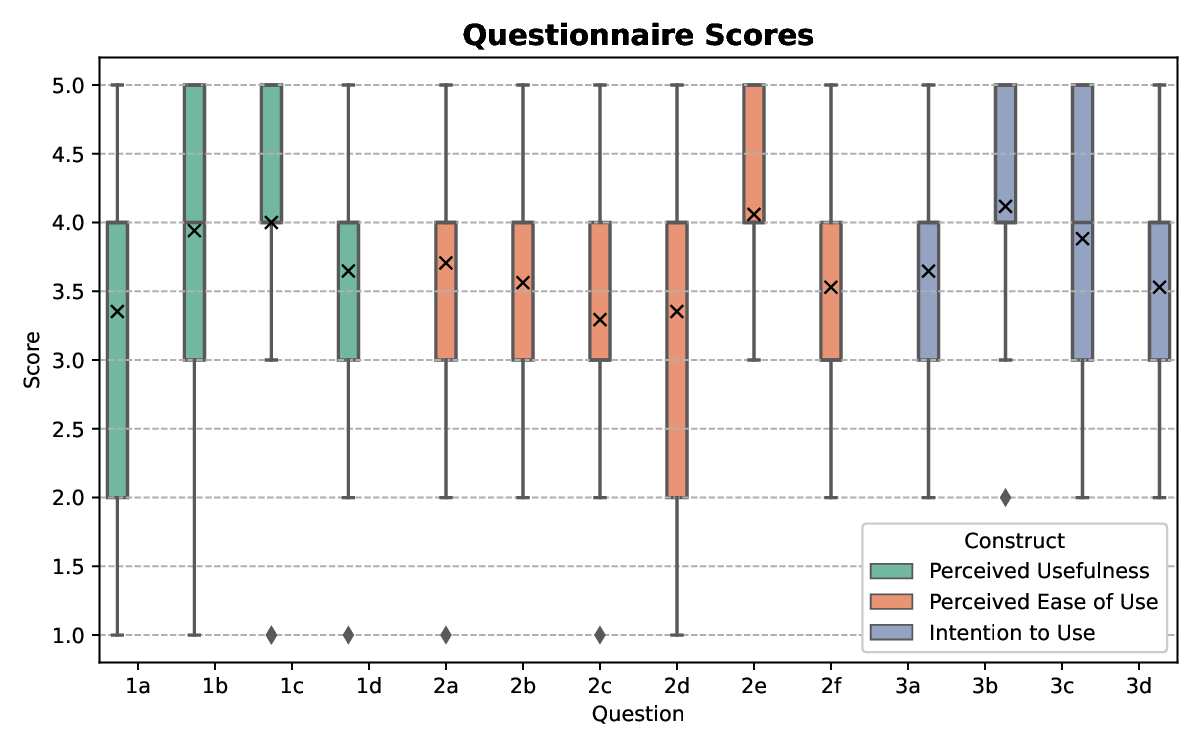}
    \caption{Questionnaire scores on a Likert scale from 1 ('Strongly Disagree') to 5 ('Strongly Agree') on the statements in \ref{app:questionnaire_statements}.}
    \label{fig:questionnaire_boxplot}
\end{figure}

Figure~\ref{fig:questionnaire_boxplot} gives an overview of the scores on the questionnaire for the different TAM constructs. Although the small sample size prevents us from drawing statistically significant conclusions, the results show that there is benefit and interest in the production trace patterns as a solution in practice. Participants indicated that they would be able to retrieve the same insights without using the patterns (question 1 of the TAM questionnaire), which shows that participants working in a variety of production environments recognise and already apply similar patterns in their data analyses. In comments at the end of the questionnaire, four participants also highlight their interest in the patterns specifically for standardising and exchanging knowledge. This further highlights the practical relevance of the work.

\section{Conclusion and Discussion}
\label{sec:conclusion}
Digital transformation has led to an increase in data collection across various systems. When data from these systems are combined and aggregated, they yield new and valuable insights. This information is often extracted ad hoc during data pre-processing. To accelerate data aggregation and facilitate the systematic exchange of knowledge regarding pre-processing procedures, we introduced generic production trace patterns. These patterns are defined using a reference model that is based on existing definitions and standards; therefore, it also serves as a basis for extracting information in settings where data are structured according to those standards.

We observed that these patterns are applicable across a variety of manufacturing settings. Practitioners working in different manufacturing domains regard them as valuable for exchanging and retaining knowledge and accelerating data aggregation. To our knowledge, this is the first work in which patterns are described for deriving hidden information from event data sets within manufacturing processes.

We recognise that restricting our study to data sets and use cases in the manufacturing field is a limitation of this work. However, this work represents an initial effort and we anticipate that the abstract patterns are applicable in other fields. For example, Pattern~\ref{pattern:interval_count} could be used in a data centre, where a computing \texttt{Resource} runs a programme to handle a job (\texttt{ProductionEntity}), with alarm \texttt{Event}s triggered by exceptions.

We do not claim that the patterns gathered here are complete. Each manufacturing setting is unique, necessitating distinct patterns to uncover hidden information. Thus, there is room for adapting and expanding patterns in diverse manufacturing environments, not limited to discrete manufacturing. Nevertheless, the collection of patterns and their implementation in SPARQL using a reference model based on standards makes them practically applicable and is a step towards systematically enriching event data. 
By providing reusable pattern definitions and template queries, this work enables the community to readily apply and extend the proposed patterns, thereby expanding the collection over time. 
The broader the pattern set, the more practitioners can benefit from new insights in their event data.

The evaluation shows the relevance of the patterns in different production environments, but should be extended to additional data sets in future. The interviews and questionnaire conducted with practitioners show that the production trace patterns are perceived as valuable, but their ease of use should be improved to support practical application. This limitation is recognised, as in this work the focus is more on the technical definition of the patterns.

Future research should therefore focus on a process and a user-friendly tool to configure the patterns so that they can be tailored to specific use cases and data sets \cite{Benevolenskiy2012ConstructionPatterns}. A promising direction to explore is the combination of a reference model with Large Language Models \cite{Shi2025EnhancingTwins}.
Future collaboration with peers could help gather and document additional patterns. Furthermore, defining validation procedures to assess and confirm the inferred information, for example, through SHACL \cite{Knublauch2017ShapesSHACL}, would be beneficial.

\section*{Acknowledgements} This work was supported by the European Union‘s Horizon 2020 research and innovation programme under grant agreement No. 957204, the project MAS4AI (Multi-Agent Systems for Pervasive Artificial Intelligence for assisting Humans in Modular Production). Additionally, it is part of the project AIMS5.0 supported by the Chips Joint Undertaking and its members, including the top-up funding by National Funding Authorities from involved countries under grant agreement no. 101112089.

%
%
\bibliographystyle{elsarticle-num-names}
\bibliography{references}

@article{Gerstl1996AApplications,
    title = {{A conceptual theory of part-whole relations and its applications}},
    year = {1996},
    journal = {Data {\&} Knowledge Engineering},
    author = {Gerstl, Peter and Pribbenow, Simone},
    number = {3},
    month = {11},
    pages = {305--322},
    volume = {20},
    publisher = {North-Holland},
    doi = {10.1016/S0169-023X(96)00014-6},
    issn = {0169-023X}
}

@article{Cao2022AManagement,
    title = {{A core reference ontology for steelmaking process knowledge modelling and information management}},
    year = {2022},
    journal = {Computers in Industry},
    author = {Cao, Qiushi and Beden, Sadeer and Beckmann, Arnold},
    month = {2},
    pages = {103574},
    volume = {135},
    publisher = {Elsevier},
    doi = {10.1016/J.COMPIND.2021.103574},
    issn = {0166-3615}
}

@phdthesis{Teymourian2014AProcessing,
    title = {{A Framework for Knowledge-Based Complex Event Processing}},
    year = {2014},
    author = {Teymourian, Kia},
    month = {11},
    school = {Freie Universit{\"{a}}t Berlin},
    address = {Berlin},
    doi = {10.17169/REFUBIUM-11103},
    keywords = {Knowledge-Based Complex Event Processing{\\\}Semantic Complex Event Processing}
}

@book{Alexander1977AConstruction,
    title = {{A Pattern Language : Towns, Buildings, Construction}},
    year = {1977},
    author = {Alexander, Christopher and Ishikawa, Sara and Silverstein, Murray},
    edition = {},
    pages = {},
    publisher = {Oxford University Press}
}

@article{Cheng2013AStreaming,
    title = {{A review of lot streaming}},
    year = {2013},
    journal = {International Journal of Production Research},
    author = {Cheng, M. and Mukherjee, N. J. and Sarin, S. C.},
    number = {23-24},
    month = {11},
    pages = {7023--7046},
    volume = {51},
    publisher = {Taylor {\&} Francis},
    doi = {10.1080/00207543.2013.774506},
    issn = {00207543}
}

@inproceedings{Alvanou2018AnAnalytics,
    title = {{An MTConnect Ontology for Semantic Industrial Machine Sensor Analytics}},
    year = {2018},
    booktitle = {SWeTI: Semantic Web of Things for Industry 4.0},
    author = {Alvanou, G. and Lytra, I. and Petersen, N.},
    pages = {57--80},
    address = {Heraklion}
}

@article{Guo2024AnProducts,
    title = {{An ontology-based method for knowledge reuse in the design for maintenance of complex products}},
    year = {2024},
    journal = {Computers in Industry},
    author = {Guo, Ziyue and Zhou, Dong and Yu, Dequan and Zhou, Qidi and Wu, Hongduo and Hao, Aimin},
    month = {10},
    pages = {104124},
    volume = {161},
    doi = {10.1016/j.compind.2024.104124},
    issn = {01663615}
}

@book{Fowler1997AnalysisModels,
    title = {{Analysis patterns: reusable object models}},
    year = {1997},
    author = {Fowler, Martin},
    edition = {},
    publisher = {Addison-Wesley},
    address = {Massachusetts},
    isbn = {0-201-89542-0}
}

@inproceedings{Muller2016CaseReasoning,
    title = {{Case Completion of Workflows for Process-Oriented Case-Based Reasoning}},
    year = {2016},
    booktitle = { Case-Based Reasoning Research and Development: 24th International Conference, ICCBR 2016},
    author = {M{\"{u}}ller, Gilbert and Bergmann, Ralph},
    pages = {295--310},
    publisher = {Springer International Publishing}
}

@inproceedings{W.Long2008ConstructOWL,
    title = {{Construct MES Ontology with OWL}},
    year = {2008},
    booktitle = {ISECS International Colloquium on Computing, Communication, Control, and Management},
    author = {{W. Long}},
    pages = {614--617},
    publisher = {IEEE},
    address = {Guangzhou}
}

@article{Benevolenskiy2012ConstructionPatterns,
    title = {{Construction processes configuration using process patterns}},
    year = {2012},
    journal = {Advanced Engineering Informatics},
    author = {Benevolenskiy, A. and Roos, K. and Katranuschkov, P. and Scherer, R. J.},
    number = {4},
    month = {10},
    pages = {727--736},
    volume = {26},
    publisher = {Elsevier},
    doi = {10.1016/J.AEI.2012.04.003},
    issn = {1474-0346}
}

@article{Giustozzi2018ContextProposal,
    title = {{Context modeling for industry 4.0: An ontology-based proposal}},
    year = {2018},
    journal = {Procedia Computer Science},
    author = {Giustozzi, F. and Saunier, J. and Zanni-Merk, C.},
    pages = {675--684},
    volume = {126}
}

@article{Li2025DataStrategy,
    title = {{Data issues in industrial AI systems: A meta-review and research strategy}},
    year = {2025},
    journal = {Computers in Industry},
    author = {Li, Xuejiao and Cheng, Yang and M{\o}ller, Charles and Lee, Jay},
    month = {12},
    pages = {104361},
    volume = {173},
    publisher = {Elsevier},
    url = {https://www.sciencedirect.com/science/article/pii/S0166361525001265#sec0045},
    doi = {10.1016/J.COMPIND.2025.104361},
    issn = {0166-3615},
    arxivId = {2406.15784}
}

@misc{DublinCore2023DCMI:Of,
    title = {{DCMI: Is Part Of}},
    year = {2023},
    author = {{Dublin Core}},
    month = {10},
    url = {https://www.dublincore.org/specifications/dublin-core/dcmi-terms/terms/isPartOf/}
}

@book{Gamma1994DesignSoftware,
    title = {{Design Patterns: Elements of Reusable Object-Oriented Software}},
    year = {1994},
    author = {Gamma, Erich and Helm, Richard and Johnson, Ralph and Vlissides, John M.},
    month = {11},
    publisher = {Addison-Wesley Professional},
    isbn = {0201633612}
}

@article{Byun2017EfficientEvents,
    title = {{Efficient and privacy-enhanced object traceability based on unified and linked EPCIS events}},
    year = {2017},
    journal = {Computers in Industry},
    author = {Byun, Jaewook and Woo, Sungpil and Kim, Daeyoung},
    month = {8},
    pages = {35--49},
    volume = {89},
    publisher = {Elsevier},
    doi = {10.1016/J.COMPIND.2017.04.001},
    issn = {0166-3615}
}

@article{Shi2025EnhancingTwins,
    title = {{Enhancing retrieval-augmented generation for interoperable industrial knowledge representation and inference toward cognitive digital twins}},
    year = {2025},
    journal = {Computers in Industry},
    author = {Shi, Dachuan and Li, Jianzhang and Meyer, Olga and Bauernhansl, Thomas},
    month = {10},
    pages = {104330},
    volume = {171},
    doi = {10.1016/j.compind.2025.104330},
    issn = {01663615}
}

@inproceedings{Anicic2011EP-SPARQL:Reasoning,
    title = {{EP-SPARQL: A Unified Language for Event Processing and Stream Reasoning}},
    year = {2011},
    booktitle = {Proceedings of the 20th international conference on World wide web},
    author = {Anicic, Darko and Fodor, Paul and Rudolph, Sebastian and Stojanovic, Nenad},
    pages = {635--644},
    isbn = {9781450306324}
}

@techreport{GS12022EPCISStandard,
    title = {{EPCIS Standard}},
    year = {2022},
    author = {{GS1}},
    month = {6},
    url = {https://ref.gs1.org/standards/epcis/},
    institution = {GS1}
}

@article{Suriadi2017EventLogs,
    title = {{Event log imperfection patterns for process mining: Towards a systematic approach to cleaning event logs}},
    year = {2017},
    journal = {Information Systems},
    author = {Suriadi, S. and Andrews, R. and ter Hofstede, A. H.M. and Wynn, M. T.},
    month = {3},
    pages = {132--150},
    volume = {64},
    publisher = {Pergamon},
    doi = {10.1016/J.IS.2016.07.011},
    issn = {0306-4379}
}

@incollection{Borgo2007FoundationsManufacturing,
    title = {{Foundations for a Core Ontology of Manufacturing}},
    year = {2007},
    booktitle = {Ontologies: a Handbook of Principles, Concepts and Applications in Information Systems},
    author = {Borgo, S. and Leit{\~{a}}o, P.},
    editor = {Sharman, Raj and Kishore, Rajiv and Ramesh, Ram},
    edition = {},
    pages = {751--775},
    volume = {14},
    publisher = {Springer},
    address = {Boston, MA}
}

@article{Afrin2025IndustrialIndustries,
    title = {{Industrial Internet of Things: Implementations, challenges, and potential solutions across various industries}},
    year = {2025},
    journal = {Computers in Industry},
    author = {Afrin, Shaila and Rafa, Sabiha Jannat and Kabir, Maliha and Farah, Tasfia and Alam, Md Sakib Bin and Lameesa, Aiman and Ahmed, Shams Forruque and Gandomi, Amir H.},
    month = {9},
    pages = {104317},
    volume = {170},
    publisher = {Elsevier},
    url = {https://www.sciencedirect.com/science/article/pii/S016636152500082X#sec0015},
    doi = {10.1016/J.COMPIND.2025.104317},
    issn = {0166-3615}
}

@inproceedings{Swevels2023InferringGraphs,
    title = {{Inferring Missing Entity Identifiers from Context Using Event Knowledge Graphs}},
    year = {2023},
    booktitle = {International Conference on Business Process Management},
    author = {Swevels, Ava and Dijkman, Remco and Fahland, Dirk},
    month = {9},
    pages = {180--197},
    publisher = {Springer Nature Switzerland},
    address = {Cham},
    isbn = {9783031416194},
    doi = {10.1007/978-3-031-41620-0{\_}11/TABLES/3},
    issn = {16113349}
}

@article{Fahland2021InferringQueues,
    title = {{Inferring Unobserved Events in Systems with Shared Resources and Queues}},
    year = {2021},
    journal = {Fundamenta Informaticae},
    author = {Fahland, Dirk and Denisov, Vadim and Van Der Aalst, Wil M P},
    number = {4},
    pages = {203--242},
    volume = {183},
    doi = {10.3233/FI-2021-2087},
    arxivId = {2103.00167v3}
}

@article{Dogan2021MachineManufacturing,
    title = {{Machine learning and data mining in manufacturing}},
    year = {2021},
    journal = {Expert Systems with Applications},
    author = {Dogan, Alican and Birant, Derya},
    month = {3},
    pages = {114060},
    volume = {166},
    publisher = {Pergamon},
    doi = {10.1016/J.ESWA.2020.114060},
    issn = {0957-4174}
}

@inproceedings{Lemaignan2006MASON:Domain,
    title = {{MASON: A proposal for an ontology of manufacturing domain}},
    year = {2006},
    booktitle = {IEEE Workshop on Distributed Intelligent Systems: Collective Intelligence and Its Applications (DIS'06)},
    author = {Lemaignan, S. and Siadat, A. and Dantan, J.Y. and Semenenko, A..},
    pages = {195–200},
    address = {Prague}
}

@article{Esser2021Multi-DimensionalDatabases,
    title = {{Multi-Dimensional Event Data in Graph Databases}},
    year = {2021},
    journal = {Journal on Data Semantics},
    author = {Esser, Stefan and Fahland, Dirk},
    number = {1-2},
    pages = {109--141},
    volume = {10},
    publisher = {Springer Science and Business Media Deutschland GmbH},
    doi = {10.1007/s13740-021-00122-1},
    issn = {18612040},
    arxivId = {18612032}
}

@article{Yang2023Ontology-basedWorkflow,
    title = {{Ontology-based knowledge representation of industrial production workflow}},
    year = {2023},
    journal = {Advanced Engineering Informatics},
    author = {Yang, Chao and Zheng, Yuan and Tu, Xinyi and Ala-Laurinaho, Riku and Autiosalo, Juuso and Sepp{\"{a}}nen, Olli and Tammi, Kari},
    month = {10},
    pages = {102185},
    volume = {58},
    publisher = {Elsevier},
    doi = {10.1016/J.AEI.2023.102185},
    issn = {1474-0346}
}

@misc{OPCFoundation2019OPCReference,
    title = {{OPC UA Online Reference}},
    year = {2019},
    author = {{OPC Foundation}},
    url = {https://reference.opcfoundation.org/v104/ISA-95/docs/4.2.3/}
}

@article{Park2025OperationalApproach,
    title = {{Operational process monitoring: An object-centric approach}},
    year = {2025},
    journal = {Computers in Industry},
    author = {Park, Gyunam and van der Aalst, Wil M.P.},
    month = {1},
    pages = {104170},
    volume = {164},
    publisher = {Elsevier},
    url = {https://www.sciencedirect.com/science/article/pii/S0166361524000988},
    doi = {10.1016/J.COMPIND.2024.104170},
    issn = {0166-3615}
}

@book{Fowler2002PatternsArchitecture,
    title = {{Patterns of Enterprise Application Architecture}},
    year = {2002},
    author = {Fowler, Martin},
    publisher = {Addison Wesley},
    address = {},
    isbn = {0-321-12742-0}
}

@incollection{Fahland2022ProcessGraphs,
    title = {{Process Mining over Multiple Behavioral Dimensions with Event Knowledge Graphs}},
    year = {2022},
    booktitle = {Process Mining Handbook},
    author = {Fahland, Dirk},
    pages = {274--319},
    volume = {448},
    publisher = {Springer Science and Business Media Deutschland GmbH},
    doi = {10.1007/978-3-031-08848-3{\_}9},
    issn = {18651356}
}

@article{Schuitemaker2020ProductReview,
    title = {{Product traceability in manufacturing: A technical review}},
    year = {2020},
    journal = {Procedia CIRP},
    author = {Schuitemaker, Reuben and Xu, Xun},
    pages = {700--705},
    volume = {93},
    publisher = {Elsevier B.V.},
    doi = {10.1016/J.PROCIR.2020.04.078},
    issn = {22128271}
}

@article{Lee2008RFID-basedChain,
    title = {{RFID-based traceability in the supply chain}},
    year = {2008},
    journal = {Industrial Management and Data Systems},
    author = {Lee, Dongmyung and Park, Jinwoo},
    number = {6},
    pages = {713--725},
    volume = {108},
    publisher = {Emerald Group Publishing Limited},
    doi = {10.1108/02635570810883978/FULL/PDF},
    issn = {02635577}
}

@misc{Knublauch2017ShapesSHACL,
    title = {{Shapes Constraint Language (SHACL)}},
    year = {2017},
    booktitle = {https://www.w3.org/TR/2017/REC-shacl-20170720/},
    author = {Knublauch, Holger and Kontokostas, Dimitris},
    url = {https://www.w3.org/TR/2017/REC-shacl-20170720/}
}

@article{Lin2015SimulationManufacturing,
    title = {{Simulation optimization approach for hybrid flow shop scheduling problem in semiconductor back-end manufacturing}},
    year = {2015},
    journal = {Simulation Modelling Practice and Theory},
    author = {Lin, James T. and Chen, Chien Ming},
    month = {2},
    pages = {100--114},
    volume = {51},
    publisher = {Elsevier},
    doi = {10.1016/J.SIMPAT.2014.10.008},
    issn = {1569-190X}
}

@misc{TheW3CSPARQLWorkingGroup2013SPARQLOverview,
    title = {{SPARQL 1.1 Overview}},
    year = {2013},
    booktitle = {https://www.w3.org/TR/2013/REC-sparql11-overview-20130321/},
    author = {{The W3C SPARQL Working Group}},
    url = {https://www.w3.org/TR/2013/REC-sparql11-overview-20130321/}
}

@article{Bonte2018StreamingStreams,
    title = {{Streaming MASSIF: Cascading Reasoning for Efficient Processing of IoT Data Streams}},
    year = {2018},
    journal = {Sensors},
    author = {Bonte, Pieter and Tommasini, Riccardo and Valle, Emanuele Della and De Turck, Filip and Ongenae, Femke},
    number = {11},
    month = {11},
    pages = {3832},
    volume = {18},
    publisher = {Multidisciplinary Digital Publishing Institute},
    doi = {10.3390/S18113832},
    issn = {1424-8220},
    pmid = {30413104}
}

@misc{OntologyDesignPatternsODP2010Submissions:PartOfOdp,
    title = {{Submissions:PartOf - Odp}},
    year = {2010},
    author = {{Ontology Design Patterns (ODP)}},
    month = {3},
    url = {http://ontologydesignpatterns.org/wiki/Submissions:PartOf}
}

@article{Davis1989TechnologyTAM,
    title = {{Technology Acceptance Model: TAM}},
    year = {1989},
    journal = {Al-Suqri, MN, Al-Aufi, AS: Information Seeking Behavior and Technology Adoption},
    author = {Davis, F.D.},
    pages = {205--219}
}

@inproceedings{Kulvatunyou2018TheProject,
    title = {{The industrial ontologies foundry proof-of-concept project}},
    year = {2018},
    booktitle = {APMS 2018: Advances in Production Management Systems. Smart Manufacturing for Industry 4.0},
    author = {Kulvatunyou, Boonserm Serm and Wallace, Evan and Kiritsis, Dimitris and Smith, Barry and Will, Chris},
    pages = {402--409},
    volume = {536},
    publisher = {Springer New York LLC},
    isbn = {9783319997063},
    doi = {10.1007/978-3-319-99707-0{\_}50/TABLES/2},
    issn = {18684238}
}

@article{Rowley2007TheHierarchy,
    title = {{The wisdom hierarchy: representations of the DIKW hierarchy}},
    year = {2007},
    journal = {Journal of information science},
    author = {Rowley, J.},
    number = {2},
    pages = {163--180},
    volume = {33}
}

@article{Usman2013TowardsOntology,
    title = {{Towards a formal manufacturing reference ontology}},
    year = {2013},
    journal = {International Journal of Production Research},
    author = {Usman, Z. and Young, R. I. M. and Chungoora, N. and Palmer, C. and Case, K. and Harding, J. A.},
    number = {22},
    pages = {6553--6572},
    volume = {51}
}

@inproceedings{Tommasini2017TowardsProcessing,
    title = {{Towards ontology-based event processing}},
    year = {2017},
    booktitle = {International Experiences and Directions Workshop on OWL},
    author = {Tommasini, Riccardo and Bonte, Pieter and Della Valle, Emanuele and Mannens, Erik and De Turck, Filip and Ongenae, Femke},
    pages = {115--127},
    volume = {10161 LNCS},
    publisher = {Springer Verlag},
    isbn = {9783319546261},
    doi = {10.1007/978-3-319-54627-8{\_}9/TABLES/1},
    issn = {16113349}
}

@article{Lee2021UnderstandingDiscovery,
    title = {{Understanding digital transformation in advanced manufacturing and engineering: A bibliometric analysis, topic modeling and research trend discovery}},
    year = {2021},
    journal = {Advanced Engineering Informatics},
    author = {Lee, Ching Hung and Liu, Chien Liang and Trappey, Amy J.C. and Mo, John P.T. and Desouza, Kevin C.},
    month = {10},
    pages = {101428},
    volume = {50},
    publisher = {Elsevier},
    doi = {10.1016/J.AEI.2021.101428},
    issn = {1474-0346}
}

@article{Schuster2022UtilizingReview,
    title = {{Utilizing domain knowledge in data-driven process discovery: A literature review}},
    year = {2022},
    journal = {Computers in Industry},
    author = {Schuster, Daniel and van Zelst, Sebastiaan J. and van der Aalst, Wil M.P.},
    month = {5},
    pages = {103612},
    volume = {137},
    publisher = {Elsevier},
    doi = {10.1016/J.COMPIND.2022.103612},
    issn = {0166-3615}
}

@article{VanderAalst2003WorkflowPatterns,
    title = {{Workflow patterns}},
    year = {2003},
    journal = {Distributed and Parallel Databases},
    author = {Van der Aalst, W. M.P. and Ter Hofstede, A. H.M. and Kiepuszewski, B. and Barros, A. P.},
    number = {1},
    month = {7},
    pages = {5--51},
    volume = {14},
    publisher = {Springer},
    doi = {10.1023/A:1022883727209/METRICS},
    issn = {09268782}
}

\appendix

\section{User evaluation questions}
\label{app:questions}

\subsection{Semi-structured Interview}
\begin{enumerate}
    \item What is your current role and function title? Would you describe your role as primarily business-oriented, IT-oriented, or a mix of both?
    \item What type of production environment(s) do you work in?
    \begin{itemize}
        \item Automotive
        \item Contract Manufacturing
        \item Industrial Equipment
        \item Food Processing
        \item Semiconductor
        \item Other:
    \end{itemize}
    \item How many years of experience do you have working with production event data?
    \begin{itemize}
        \item 0-2
        \item 3-5
        \item 6-10
        \item 10+
    \end{itemize}
    \item What systems do you typically work with?
    \begin{itemize}
        \item ERP (Enterprise Resource Planning, e.g. SAP)
        \item MES (Manufacturing Execution System)
        \item SCADA (Supervisory Control and Data Acquisition)
        \item Sensors
        \item Other:
    \end{itemize}
    \item Which of the following best describes how you typically approach analytical questions?
    \begin{itemize}
        \item I help colleagues answering analytical questions they have.
        \item I'm hands on working with data and can answer the analytical questions I have myself.
        \item I can answer some standard analytical questions myself, but need help from colleagues to get deeper insights.
        \item I typically go to (IT) colleagues to get answers to analytical questions I have.
    \end{itemize}
    \item On average, how many analytical questions do you need to answer per week using production event data?
    \item How do you typically answer these questions?
    \begin{itemize}
        \item Manual inspection
        \item BI tools (e.g. Power BI)
        \item SQL queries
        \item Python/R scripts
        \item Custom dashboards
        \item Other:
    \end{itemize}
    \item If applicable, approximately how many lines of code (script or SQL) do you write per analytical question?
    \item If applicable, how much effort (hours of work) is typically required to answer one analytical question? 
    \item If applicable, how many working days do you typically have to wait to get your analytical questions answered?
    \item What analytical questions do you commonly ask?
    \item What are the biggest challenges you face when analyzing production event data? And/or what factors help to deal with those challenges?
\end{enumerate}

\subsection{Technology Acceptance Questionnaire}
\label{app:questionnaire_statements}
\begin{enumerate}
    \item \textbf{Perceived Usefulness}
    \begin{enumerate}
        \item The trace patterns provide insights I wouldn’t easily get otherwise.
        \item Applying trace patterns improves the quality of my data-driven decisions.
        \item The ability to aggregate, relate, and calculate time between events adds value to my data analysis tasks.
        \item I can see clear benefits of applying this solution in my daily work.
    \end{enumerate}
    \item \textbf{Perceived Ease of Use}
    \begin{enumerate}
        \item I quickly understood how to apply the reference model and trace patterns.
        \item The patterns are intuitive and easy to apply to my production data.
        \item The solution integrates well with my existing data workflows and tools.
        \item I can apply the patterns without needing extensive training or domain-specific programming skills.
        \item The visual representations of trace patterns help me understand their logic and application.
        \item Filling in the templates is straightforward and does not require deep technical expertise.
    \end{enumerate}
    \item \textbf{Behavioural Intention to Use}
    \begin{enumerate}
        \item I intend to use this solution in future production data analysis tasks.
        \item I would be willing to invest time to learn more about this solution.
        \item I would recommend this solution to colleagues working with event data.
        \item I plan to explore more trace patterns to enrich my datasets.
    \end{enumerate}
\end{enumerate}

\section{Summary of answers to semi-structured interview questions}
\label{app:descriptives}

\begin{table}[H]
    \centering
    \caption{Overview of roles of the participants (`What is your current role and function title? Would you describe your role as primarily business-oriented, IT-oriented, or a mix of both?').}
    \begin{tabular}{l|l}
        \textbf{Role (title)} & \textbf{Business- or IT-oriented} \\ \hline
        \parbox[t]{7cm}{System architect and analyst} & mix of both \\ \hline
        \parbox[t]{7cm}{Logistic Engineer} & mix of both \\ \hline
        \parbox[t]{7cm}{Quality engineer focusing on traceability data} & \parbox[t]{4cm}{business-oriented with bridge to IT} \\ \hline
        \parbox[t]{7cm}{System Analyst in Quality Information Systems} & \parbox[t]{4cm}{business-oriented with bridge to IT} \\ \hline
        \parbox[t]{7cm}{Data Scientist and Production Process Engineer} & mix of both \\ \hline
        \parbox[t]{7cm}{Founder and product responsible (solution provider)} & \parbox[t]{4cm}{business-oriented with bridge to IT} \\ \hline
        \parbox[t]{7cm}{Product architect of data platform (data analytics organization)} & IT-oriented \\ \hline
        \parbox[t]{7cm}{Center of Excellence director planning} & business-oriented \\ \hline
        \parbox[t]{7cm}{Data-driven optimization specialist} & \parbox[t]{4cm}{business-oriented with bridge to IT} \\ \hline
        \parbox[t]{7cm}{Six-sigma problem solver (master black belt)} & business-oriented \\ \hline
        \parbox[t]{7cm}{Project manager in the Operational Excellence team} & business-oriented \\ \hline
        \parbox[t]{7cm}{Business Intelligence Engineer - Business Software architect} & IT-oriented \\ \hline
        \parbox[t]{7cm}{Manager maintenance and sustainability} & business-oriented \\ \hline
        \parbox[t]{7cm}{Senior Data Scientist Supply Chain} & mix of both \\ \hline
        \parbox[t]{7cm}{Industrial Engineer} & mix of both \\ \hline
        \parbox[t]{7cm}{Senior Industrial Engineer} & mix of both
    \end{tabular}
\end{table}

\begin{table}[H]
    \centering
    \caption{Overview of answers on the multiple choice questions of the semi-structured interview describing the role of the participant.}
    \begin{tabular}{l|r|c}
        \textbf{Question}                                                                      & \textbf{Answer}                                  & \textbf{Count} \\ \hline
        \multirow{5}*{\parbox[t]{4cm}{What type of production environment(s) do you work in?}} & Automotive                                       & 1              \\
                                                                                               & Contract Manufacturing                           & 4              \\
                                                                                               & Food processing                                  & 4              \\
                                                                                               & Industrial Equipment                             & 1              \\
                                                                                               & Semiconductor                                    & 7              \\ \hline
        \multirow{4}*{\parbox[t]{4cm}{How many years of experience do you have working with production event data?}}           & 0-2              & 1              \\
                                                                                               & 3-5                                              & 4              \\
                                                                                               & 6-10                                             & 7              \\
                                                                                               & 10+                                              & 4              \\ \hline
        \multirow{5}*{\parbox[t]{4cm}{What systems do you typically work with?}}               & ERP                                              & 11             \\
                                                                                               & MES                                              & 10             \\
                                                                                               & SCADA                                            & 7              \\
                                                                                               & Sensors                                          & 4              \\
                                                                                               & Other                                            & 9              \\ \hline
        \multirow{5}*{\parbox[t]{4cm}{Which of the following best describes how you typically approach analytical questions?}} & Help colleagues  & 16             \\
                                                                                               & Hands on                                         & 14             \\
                                                                                               & Help required for complex questions              & 4              \\
                                                                                               & Go to colleagues                                 & 1              \\ 
                                                                                               &                                                  &                \\ \hline
        \multirow{6}*{\parbox[t]{4cm}{How do you typically answer these questions?}}           & Manual inspection                                & 11             \\
                                                                                               & BI tools (e.g. Power BI)                         & 7              \\
                                                                                               & SQL queries                                      & 9              \\
                                                                                               & Python/R scripts                                 & 10             \\
                                                                                               & Custom dashboards                                & 5              \\
                                                                                               & Other                                            & 5              
    \end{tabular}
\end{table}

\begin{table}[H]
    \centering
    \caption{Statistics for the answers to the numeric questions of the semi-structured interview. Note that due to the low sample size and high variability in the answers, it is not possible to draw statistically significant conclusions, but the numbers do show the expertise of the participants with data analysis and aggregation tasks.}
    \begin{tabular}{l|c|c|c}
        \textbf{Question} & \textbf{Count} & \textbf{Average} & \textbf{Standard deviation} \\ \hline
        \parbox[t]{5cm}{On average, how many analytical questions do you need to answer per week using production event data?} & 16 & 8.0 & 7.4 \\ \hline
        \parbox[t]{5cm}{If applicable, approximately how many lines of code (script or SQL) do you write per analytical question?} & 12 & 250 & 362 \\ \hline
        \parbox[t]{5cm}{If applicable, how much effort (hours of work) is typically required to answer one analytical question?} & 16 & 11 & 15 \\ \hline
        \parbox[t]{5cm}{If applicable, how many working days do you typically have to wait to get your analytical questions answered?} & 9 & 15 & 23
    \end{tabular}
\end{table}

\section{Occurrence of themes in analytic questions}
\label{app:themes_analytical_questions}

\begin{figure}[H]
    \centering
    \begin{subfigure}[b]{0.48\textwidth}
        \centering
        \includegraphics[width=\textwidth]{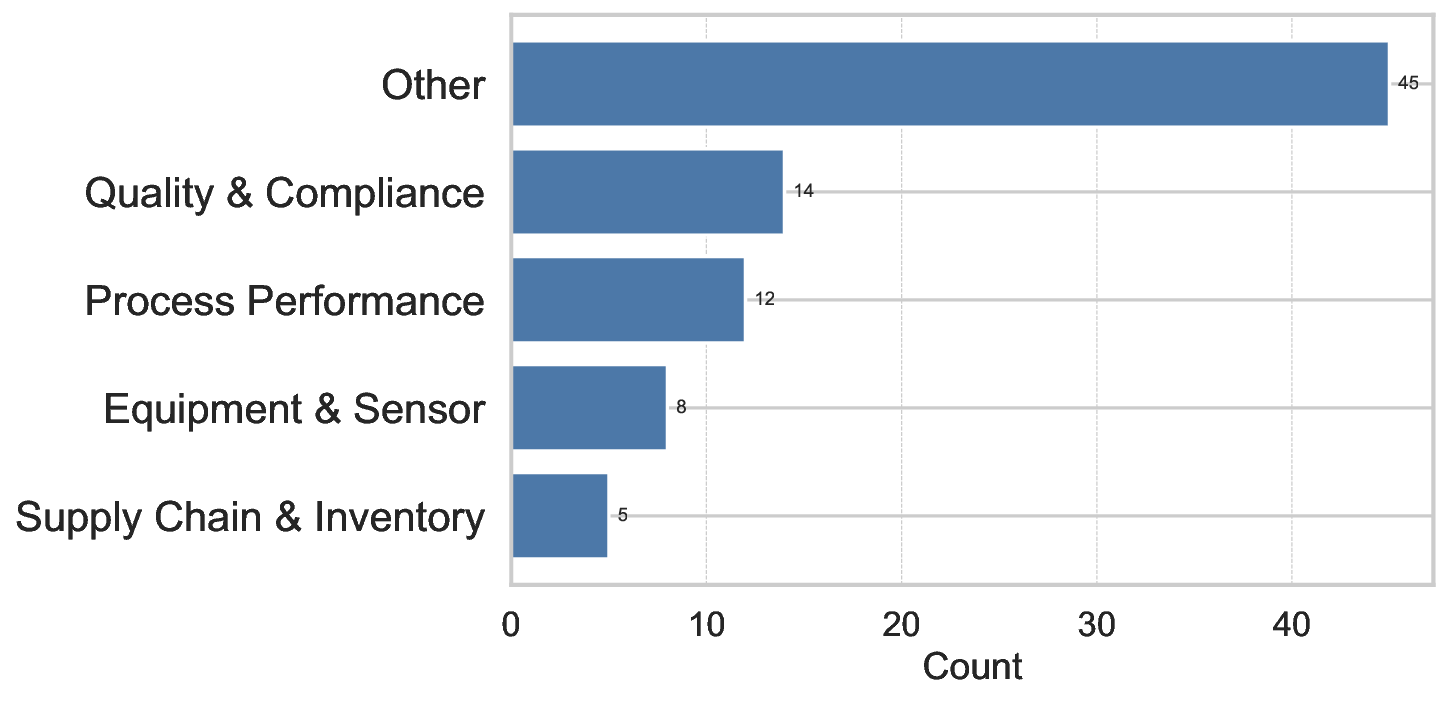}
        \caption{Occurrence of topics.}
        \label{fig:topic_themes_question11}
    \end{subfigure}
    \hfill
    \begin{subfigure}[b]{0.48\textwidth}
        \centering
        \includegraphics[width=\textwidth]{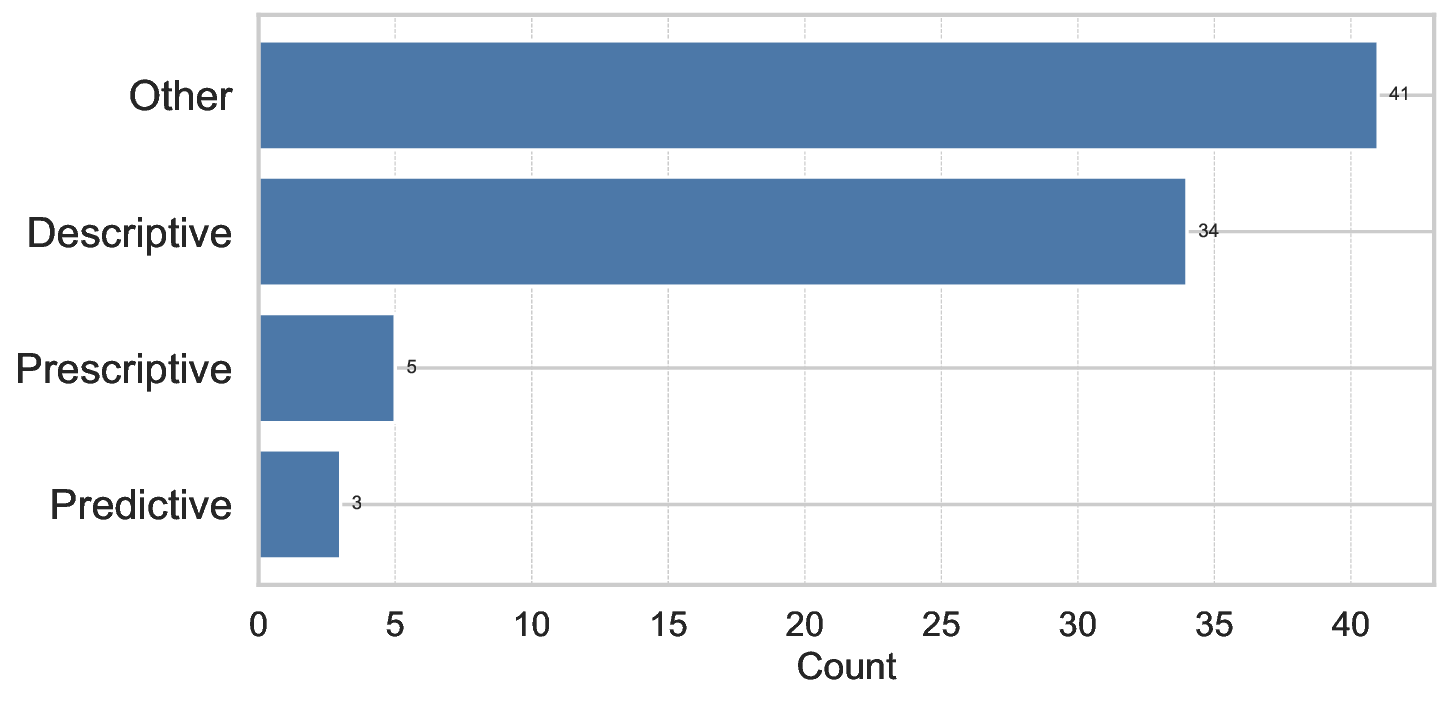}
        \caption{Occurrence of analytic types.}
        \label{fig:type_themes_question11}
    \end{subfigure}
    
    \caption{Occurrence of themes in the answers to question 11 (`What analytical questions do you commonly ask?').}
    \label{fig:themes_question11}
\end{figure}

\section{Co-occurrence of themes in analytic questions and challenges}
\label{app:cooccurrence_question_11_12}

\begin{figure}[H]
    \centering
    \includegraphics[width=0.75\linewidth]{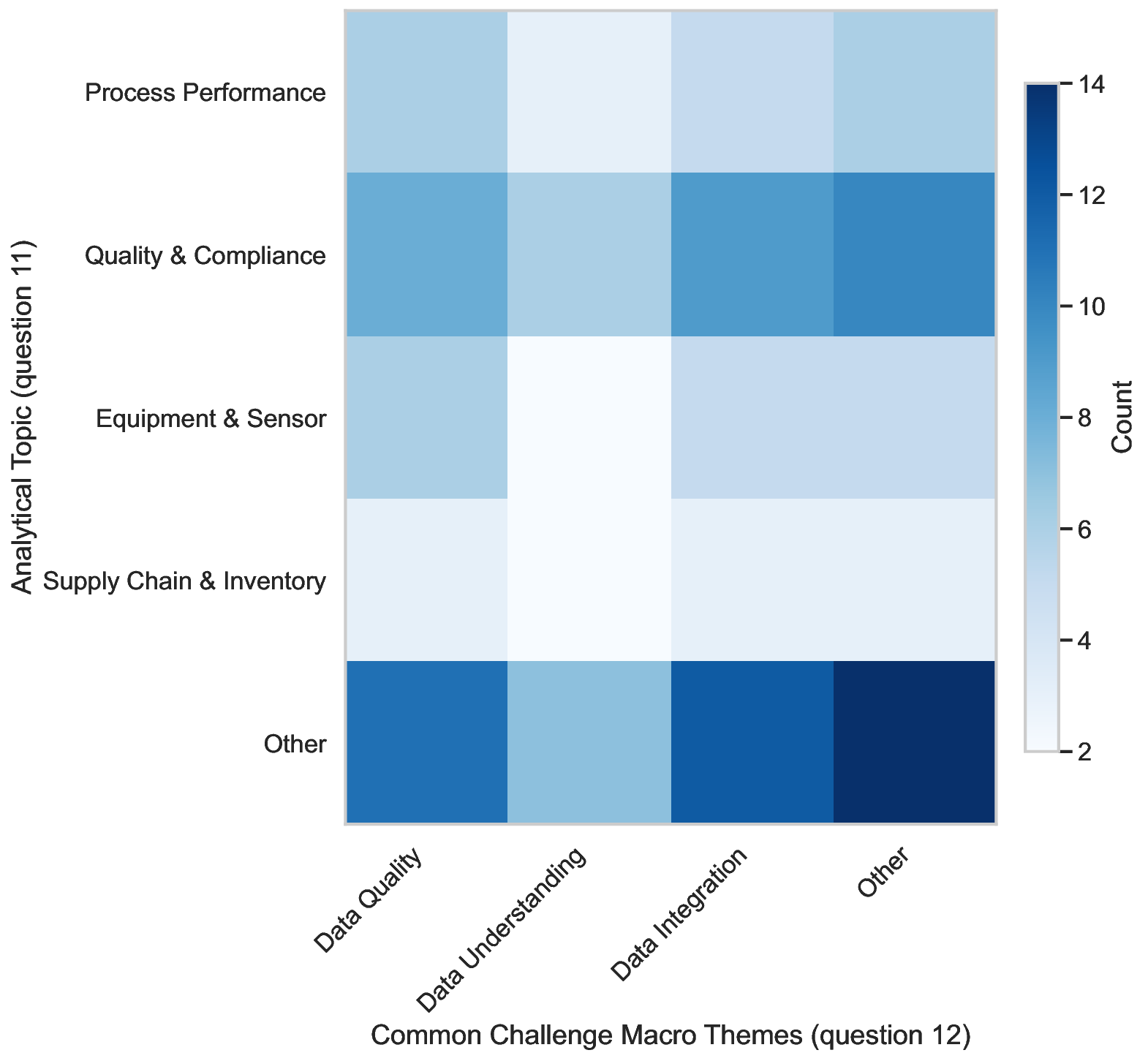}
    \caption{Heat map showing co-occurrence of themes in the answers to questions 11 (topics of analytical questions) and 12 (common challenges).}
    \label{fig:heatmap_questions_challenges}
\end{figure}

\end{document}